# Optimization and application of ultra-high field preclinical high-resolution and 3D $^1$H-MRSI using compressed sensing

Brayan Alves[1,2], Thanh Phong Lê[1,2], Gianna Nossa[1,2], Tan Toi Phan[1,2], Alessio Siviglia[1,2], Bernard Lanz[1,2], Wolfgang Bogner[3,4,5], Antoine Klauser[6], Bernhard Strasser[3], Cristina Cudalbu[1,2]

[1] CIBM Center for Biomedical Imaging, Switzerland

[2] Pre-clinical Imaging PCI, École polytechnique fédérale de Lausanne (EPFL), Lausanne, Switzerland

[3.] High-field MR Center, Department of Biomedical Imaging and Image-guided Therapy, Medical University Vienna, Vienna, Austria

[4] Christian Doppler Laboratory for MR Imaging Biomarkers (BIOMAK), Austria.

[5] Christian Doppler Laboratory for MR Imaging Biomarkers (BIOMAK), Austria.

[6] Swiss Innovation Hub, Siemens Healthineers International AG, Lausanne, Switzerland

**Abbreviations:**

UHF – Ultra-high Field, MRSI - Magnetic Resonance Spectroscopic Imaging, FID - Free Induction Decay, AD - Acquisition Delay, TR - Repetition Time, CS – Compressed Sensing, UF – Undersampling factor, LR-TGV - Low-Rank Total Generalized Variation, SVD - Singular Value Decomposition, SNR - Signal to Noise Ratio, $SNR_t$ – SNR per unit of time, SD - Standard Deviation, FWHM - Full Width at Half Maximum, NAA - N-Acetyl Aspartate, NAAG - N-acetylaspartylglutamate, Ins - myo-Inositol, Gln - Glutamine, Glu - Glutamate, GABA - Gamma-Aminobutyric Acid, Asp - Aspartic Acid, Ala - Alanine, Asc - Ascorbic Acid, Cr - Creatine, PCr - Phosphocreatine, tCr - total Creatine, Tau - Taurine, PCho - Phosphocholine, GPC - Glycerophosphocholine, PE - Phosphoethanolamine, Lac - Lactate, Glc - Glucose, GSH - Glutathione

# Abstract

Proton magnetic resonance spectroscopic imaging ($^1$H-MRSI) at ultra-high field has seen an increase in usage in the preclinical field. Challenges related to long acquisition time and low concentration of brain metabolites in the rodent brain have led to the development and application of acceleration schemes for 3D-$^1$H-MRSI, such the undersampling technique Compressed Sensing (CS).

This present study aims to explore the CS tool in the context of preclinical *in vivo* application in order to achieve high-resolution MRSI acquisition in both 2D with an in-plane increase and 3D/multi-slice acquisition with through-plane. The parameters are explored to achieve the highest acceleration possible as a way to make 3D as time efficient as possible.

Results of the parameter study showed that an acceleration factor (AF) of 4 was possible with the right sampling size of the core at the center of the *k*-space. With this specific set, higher matrix size resulting in sub 1 µL nominal voxel size was explored with 2D-FID-MRSI and 9 supplementary phase-encoding/slices were added to achieve 3D-FID-MRSI. The spectral quality and the metabolic maps were accurate enough in the comparison with the non-accelerated 2D-FID-MRSI, within the slice of interest. Issues related with the point spread function (PSF) were noted throughout the different usage of CS.

Our work presents a robust and effective protocol to achieve 3D-$^1$H-MRSI using CS in order to reach an acquisition time below the 30 minutes bar, with minimal technical limitations and high-quality acquisition.

# 1. Introduction

Proton magnetic resonance spectroscopic imaging $^1$H-MRSI is a spectroscopic technique that allows the acquisition of multiple MR spectra at different spatial locations using phase encoding. This tool provides the ability to do non-invasive *in vivo* metabolic imaging, which is valuable for region-specific metabolic analysis in different organs such as the brain[1,2]. Free induction decay MRSI (FID-MRSI) at ultra-high field (UHF) leads to faster acquisitions, has seen an increase in usage in the clinical context[3,4], and has been key in the investigation of numerous neurodegenerative pathologies[2,5,6]. Recently, $^1$H-FID-MRSI gained traction in the preclinical realm with an implementation of this sequence at 9.4T and 14.1T resulting in reliable metabolic distribution in the rodent brain[7,8]. However, limitations with regards to low signal-to-noise ratio (SNR) and long acquisition time have been mentioned in both clinical and preclinical fields[9,10]. Few solutions have been explored to deal with these limitations in the clinical context, such as noise reduction post-processing algorithms[11–13] and acceleration schemes using compressed sensing (CS) or spatial-spectral encoding[10,14–16]. The reduction in acquisition time has allowed more efficient exploration of higher coverage using multi-slice[17] or three-dimensional (3D)-MRSI[18], opening the path to more extensive brain pathologies such as glioma[19], sclerosis[20] or Alzheimer's disease[21]. Preclinical post-processing denoising tools have been tested to bypass SNR related issues[13] and acceleration techniques have found usage in the preclinical hyperpolarized X-nuclei MRSI[22–24]. However, to the authors' knowledge, no usage of acceleration techniques or its application for 3D metabolic mapping has been reported for non-hyperpolarized preclinical $^1$H-FID-MRSI.

CS is a compression technique commonly used in standard clinical MRI[25,26], and in $^1$H and X-nuclei MRSI[27–29]. This technique allows accelerated MRSI data acquisition by employing sparse *k*-space sampling combined with a dedicated reconstruction[30]. Due to its ease of application, CS can be used for multislice or 3D-MRSI by either applying the undersampling on each slice independently or on the three dimensions uniformly, respectively. The acceleration factor (AF) is defined as the inverse of the fraction of the sampled *k*-space. The advantages of CS-MRSI are that it is not subjected to *g*-factor penalty, does not require calibration scans, and can be combined with advanced reconstruction methods such as LowRank to achieve high-resolution metabolite imaging[16,27]. Despite these advantages, CS has not yet been used for $^1$H-FID-MRSI in preclinical studies.

The aim of the present study was to implement an optimized protocol for preclinical usage of CS $^1$H-FID-MRSI at ultra-high field, in the rodent brain with the final aim of decreasing the acquisition time while increasing the brain coverage, e.g., by going to 3D-MRSI. The evaluation of the protocols were conducted by estimating the negative impact of CS k-space sampling on the spatial resolution (e.g., point spread function, PSF) and the appearance of acceleration related artifacts in the form of lipid signal contamination or noise-like aliasing[31] on the spectra relative to the gains in temporal resolution.

These protocols were then used for acquisition of in-plane higher-resolution $^1$H-FID-MRSI (sub- 1 µL nominal voxel size acquisition), for multi-slice, and 3D $^1$H-FID-MRSI. Each of these methodologies were assessed in their abilities to generate reliable metabolic maps with sufficient coverage and SNR, as a way to motivate faster acquisition of preclinical $^1$H-FID-MRSI at UHF while preserving spectral quality.

# 2. Methods

*In vivo* acquisitions conducted in this study were approved by The Committee on Animal Experimentation for the Canton de Vaud, Switzerland (VD 3892). Wistar adult rats (*n* = 16 rats, 253 ± 50 g, Charles River Laboratories, L'Arbresle, France) under 1.5-2.5% isoflurane anesthesia were used. The body temperature of the animals was kept at 37.5 ± 1.0 °C by circulating warm water and measured with a rectal thermosensor. The respiration rate and body temperature were monitored using a small-animal monitor system (SA Instruments, New York, NY, USA). During the acquisition of MRI and MRSI data, all animals were placed in an in-house-built or in a Bruker-manufactured holder, with their head fixed in a stereotaxic system using a bite bar and a pair of ear bars.

CS optimization for $^1$H-FID-MRSI measurements was performed in the rat brain (*n* = 8 rats) on a 14.1T horizontal magnet (Magnex Scientific, Yarnton, UK), with a 1 T/m peak strength and 5500 T/m/s slew rate shielded gradient set (Resonance Research, Billerica, USA) interfaced to a Bruker console (BioSpec AVANCE NEO, ParaVision 360 v3.3), and using a home-made transmit/receive quadrature surface coil (1.8 × 1.6 cm inner loop radius). 2D High-resolution, multi-slice and 3D $^1$H-FID-MRSI acquisitions (*n* = 8 rats) were acquired on a 9.4T horizontal magnet (Magnex Scientific, Yarnton, UK), with a 660 mT/m peak strength and 4570 T/m/s slew rate shielded gradient set (Bruker B-GA12S HP) interfaced to a Bruker console (BioSpec AVANCE NEO, ParaVision 360 v3.5), and using a quadrature volume-transmit coil and a cryogenic four-channel receive array for the rat head (CryoProbe, Bruker).

Coronal and axial MRI images were acquired on both systems using a T$_2$-weighted Turbo-RARE images for positioning of the MRSI slice, brain segmentation and shimming (TR = 4100 ms, TE$_{eff}$ = 27 ms, NA = 10, RARE$_{factor}$ = 6, 128 × 128 matrix, FOV = 24 × 24 mm). The number of slices and their thickness were adapted for each system: 60 slices of 0.2 mm thickness for 14.1T and 40 slices of 0.3 mm thickness for 9.4T. The shimming procedure for preclinical 2D $^1$H-FID-MRSI described by Simicic et al.[7] and found in [LIVE Demos – MRS4BRAIN - EPFL](#) was used throughout this study. The same procedure has been adapted for multi-slice and 3D measurements by increasing the shimming volume (from 10 × 10 × 2 mm to 10 × 10 × 6 mm) to match the MRSI slab thickness used.

## 2.1. CS-MRSI Sequence

### 2.1.1. CS optimization at 14.1T

2D $^1$H-MRSI data for exploration of CS parameters were acquired in the rat brain in a region covering the hippocampus and a part of the striatum, using the recently implemented preclinical single slice fast $^1$H-FID-MRSI sequence and protocol[7]. MRSI was performed with one average only and a pair of water and metabolite acquisition was acquired for both fully and under-sampled acquisition.

2D cartesian *k*-space sampling was used with two phase encoding gradients (Supplementary Figure 1, left-side): the *RAW* MRSI data which was fully sampled (i.e. 961 points, 1 average, 13 min) and the *CS* MRSI data which was under-sampled using CS. For the CS MRSI acquisition, two parameters were used to proceed with the under-sampling during the acquisition: the percentage of volume sampled (referred in this manuscript as *UF*) and the percentage of the overall volume that is fully sampled at the center of the *k*-space (referred in this manuscript as *Core*). *UF* describes the number of *k*-space points acquired during a CS acquisition ($961 \times UF \, 10^{-2}$) as well as AF($= \frac{100}{UF}$). Consequently, the acquisition time of a CS procedure is defined by $\frac{13}{AF}$ minutes. *Core* describes how much of the *k*-space is completely sampled. This implies that the maximum value *Core* can reach in a CS acquisition is *UF*. The core fully sampled is located at the center of the *k*-space, in the form of a rectangle. All the *k*-space points that fall outside the core are sampled randomly, with a uniform distribution (Supplementary Figure 1, right-side). In this study, tests on the AF factor were done with a fixed *Core* = 20% while tests on the *Core* were conducted with AF = 2. As the undersampling was only applied in the spatial dimension, it was also assumed that the optimized configurations would work on both 14.1T and 9.4T systems. To respect the 3R conventions, it was decided that the optimization would be done once on only one system (14.1T).

### 2.1.2. 2D-High Resolution CS-FID-MRSI at 9.4T

Due to the possibility to acquire data with a CryoProbe, the 2D-FID-MRSI sequence was adapted to fit acquisitions at the 9.4T system by using a 5 kHz acquisition bandwidth and 768 spectral data points. The number of saturation bands was increased from 7 to 12 to avoid lipid contamination due to the increase in excitation coverage caused by the volume-transmit coil relative to the surface-transmit used at the 14.1T. Consequently, the TR increased from 813 ms to 822 ms and the flip angle was adjusted accordingly to 55.1°[32]. The Shinnar-Le Roux optimized pulse was preserved, leading to AD = 1.3 ms. The positioning procedure and the geometry of the MRSI slice was kept the same.

Each $^1$H-FID-MRSI sets were acquired with three different nominal voxel size by changing the matrix size: *31 × 31* resulting in 0.77 × 0.77 × 2 mm (1.19 µL), *47 × 47* resulting in 0.51 × 0.51 × 2 mm (0.52 µL) and *63 × 63* resulting in 0.38 × 0.38 × 2 mm (0.29 µL). CS was used with an AF of 4 and an optimized *Core* of 20% to reduce the increase in acquisition time for *47 × 47* and *63 × 63*. Moreover, to compensate for the loss in the SNR, measurements were done with 3 averages for the two high-resolution acquisitions, leading to an acquisition time of 23 and 41 minutes for *47 × 47* and *63 × 63* respectively, instead of 91 and 163 minutes without CS. All the specific parameters are reported in Supplementary Table 2.

### 2.1.3. Multi-Slice and 3D CS-FID-MRSI at 9.4T

Multi-slice and 3D FID-MRSI slabs were acquired in the brain with one slice encoded at the center of the 2D FID-MRSI acquired slice (named *2D* in this section and for further comparison in the manuscript). The FOV was extended to 24 × 24 × 9 mm for a matrix resolution of 31 × 31 × 9, implying 9 slices of 1 mm thickness in the slice encoding direction. Because of the larger excitation volume, an additional saturation slab, positioned on top of the cranium of the rat, was added to the 12 defined in the 2D sequence. The positioning of the saturation slices can be observed in Figure 1A.

For Multi-slice acquisitions (*Multislice*), the 2D's original thickness was reduced to 1 mm and repeated 9 times for each slice selection. This methodology works sequentially: each phase encoded in-plane point is acquired on each slice individually before moving to the next. This leads to a multiplication of the TR ($TR_{Multislice}$ = 9 × 822 ms) and a correction of the flip angle to 90°. CS was thus applied per slice, with an AF = 4 and a *Core* = 20%. Consequently, the effective *k*-space center acquired takes the shape of a parallelepiped whose length is on the slice selection direction. A sketch of the Multislice sequence and the undersampling pattern is displayed on the left side of Figure 1B.

For 3D FID-MRSI (*3D*), the parameters used for 2D were preserved with an additional phase encoding applied with the slice selective gradient. Due to the anisotropic nature of the excitation slab, the *Core* parameter of CS was adapted automatically by the vendor acquisition sequence from 20% to 3%, while preserving the AF at 4. With this change, the effective *k*-space pattern was more uniformly distributed on the volume of interest (right side of Figure 1B). The acquisition time for both Multislice and 3D was shortened from 118 min to approximately 29 min and the nominal voxel size, compared to 2D, was reduced from 1.19 µL to 0.59 µL.

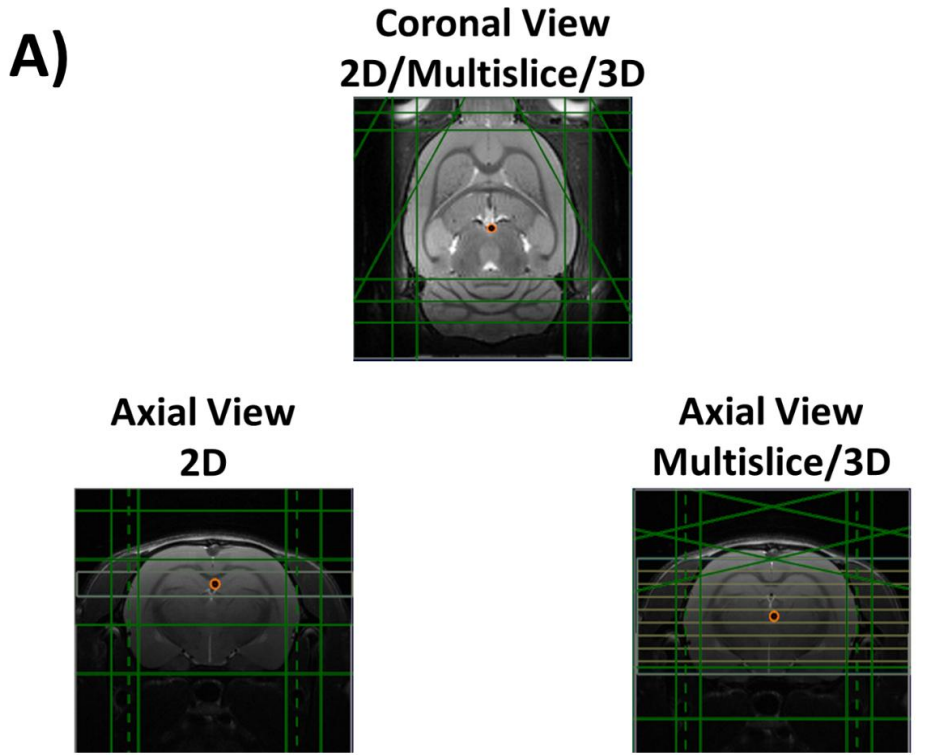
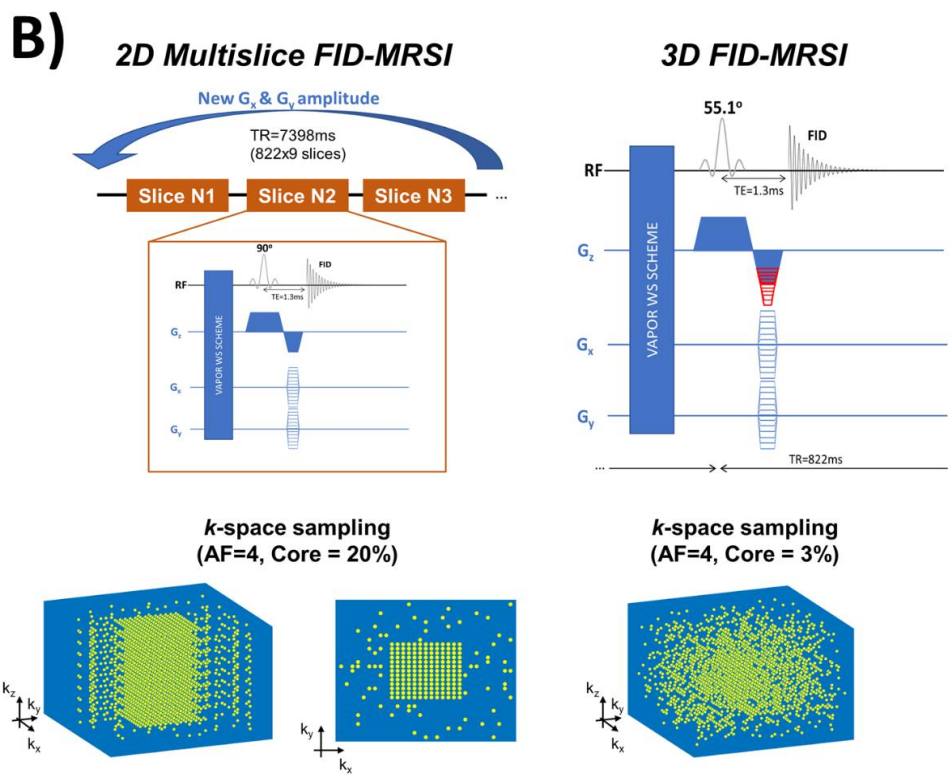

*Figure 1 A) FOV saturation slice positions from the coronal and axial view of 2D, Multislice and 3D. The main differences are noted from the axial view: the FOV saturation slices at the lower position was pushed lower to accommodate the bigger excitation slice and the additional saturation slice at the top was placed in a way to create a "rooftop" configuration on the cranium. B) Sketch of the 3D and Multislice FID-MRSI sequence, as well as the k-space sampling modes used for both respectively*

### 2.2.3. Processing and Fitting Procedure

An in-plane 2D and 3D *k*-space Hamming filter (built in Paravision V.3.3 and V.3.5) was applied on each spatial direction. The filter was applied directly on all acquired MRSI dataset in the *k*-space (before reconstruction).

All $^1$H-FID-MRSI data acquired in this study were processed using the *MRS4Brain Toolbox*[7,33]. The additional water acquisition was used for filtering out the voxels outside the brain region with a power mask and for quality control (QC) via the water linewidth and the $B_0$-shift distribution. HLSVD water removal algorithm was used for residual water signal removal. The datasets were quantified using LCModel (Version 6.3-1N). Two metabolite basis-sets containing 18 metabolites (alanine (Ala), aspartate (Asp), ascorbate (Asc), creatine (Cr), phosphocreatine (PCr), γ-aminobutyrate (GABA), glutamine (Gln), glutamate (Glu), glycerophosphorylcholine (GPC), glutathione (GSH), glucose (Glc), inositol (Ins), N-acetylaspartate (NAA), N-acetylaspartylglutamate (NAAG), phosphorylcholine (PCho), phosphorylethanolamine (PE), lactate (Lac), and taurine (Tau)) were simulated, following the procedure described by Simicic et al.[7] using published J-coupling and chemical shift values[34,35], for the different magnet intensity using NMRScopeB[34–36], with the same pulse-acquire protocol as used for *in vivo* acquisitions. The macromolecules were acquired *in vivo* using a double inversion recovery module, combined with the FID-MRSI sequence[37]. Seven and 2 voxels were summed to obtain the final MM signal, at 14.1T and 9.4T, respectively. Both water and metabolite residuals were removed from the macromolecules spectrum using AMARES as previously described[38,39] (more information is available in: LIVE Demos – MRS4BRAIN - EPFL). PCho and GPC, and Cr and PCr were expressed only as tCho (PCho + GPC) and tCr (Cr + PCr) due to better accuracy in their estimation as a sum (more informations on Supplementary Table 1), and tCr was used as internal reference. Due to the difference in TR between 2D/3D and Multislice, a T1 correction factor was applied on the estimates of the Multislice acquisition.

A semi-automatic QC filter implemented in the *MRS4Brain Toolbox* was applied, using the SNR and linewidth (measured with Full Width at Half Maximum, FWHM) metrics. These values are obtained from the LCModel outputs, and the thresholds are set at minimum SNR = 4[40,41] and maximum FWHM values at 125% of the average over the number of voxels[42]. For each metabolite, a specific mask with respect to the Cramer−Rao lower bound (CRLB) values was used (≤ 30%).

### 2.3. Data Display, Metrics and Statistics

An atlas-based automatic segmentation using the coronal MRI images was performed with the *MRS4Brain Toolbox*. At each magnetic field, a homemade template aligned with the SIGMA atlas[43] (see Supplementary Material 3) was used to depict the hippocampus, dorsal striatum, and cortex (Figure 2). Relative concentrations are presented as mean ± standard deviation (SD) across regions and afterwards datasets (i.e., number of rats). The two-way analysis of variance (ANOVA), available in the toolbox, with respect to each metabolite in the neurochemical profile, followed by Bonferroni's multi-comparisons post-hoc were performed. Two categorical factors were defined for the different CS applications (31 × 31, 47 × 47, 63 × 63 / 2D, Multislice, 3D) and for the brain regions (Hippocampus and Cortex + Striatum). The significance level was attributed as follows: *$p < 0.05$, **$p < 0.01$, ***$p < 0.001$, and ****$p < 0.0001$. All tests were two-tailed.

Additional quality metrics were computed to evaluate objectively the efficiency of the CS datasets and its application on higher resolution acquisition. For parameter optimization, the SNR per unit of time ($SNR_t$) was computed by dividing the SNR averaged over the number of voxels by the square root of the overall acquisition time. For comparison between the different resolutions, the normalized SNR defined as the $SNR_t$ divided by the voxel volume was used. Between the different reconstruction types, the coefficients of variation maps, calculated as the per voxel ratio of one SD over the mean of the relative concentration estimated between the sets, were computed. The pre-reconstruction PSF (for the sake of clarity, it will preserve its abbreviation: PSF) was also estimated each time when CS was applied, to estimate the various spatial-related penalties. Three main metrics were analyzed in this regard: FWHM of the PSF, the maximum lobe amplitude, and its approximate propagation in the number of voxels. The lobe amplitude was estimated by assessing the highest value of the absolute profile of the PSF outside of its central lobe. The effective voxel size was also calculated from the resulting PSF by multiplying the length of the nominal voxel size by the FWHM estimate.

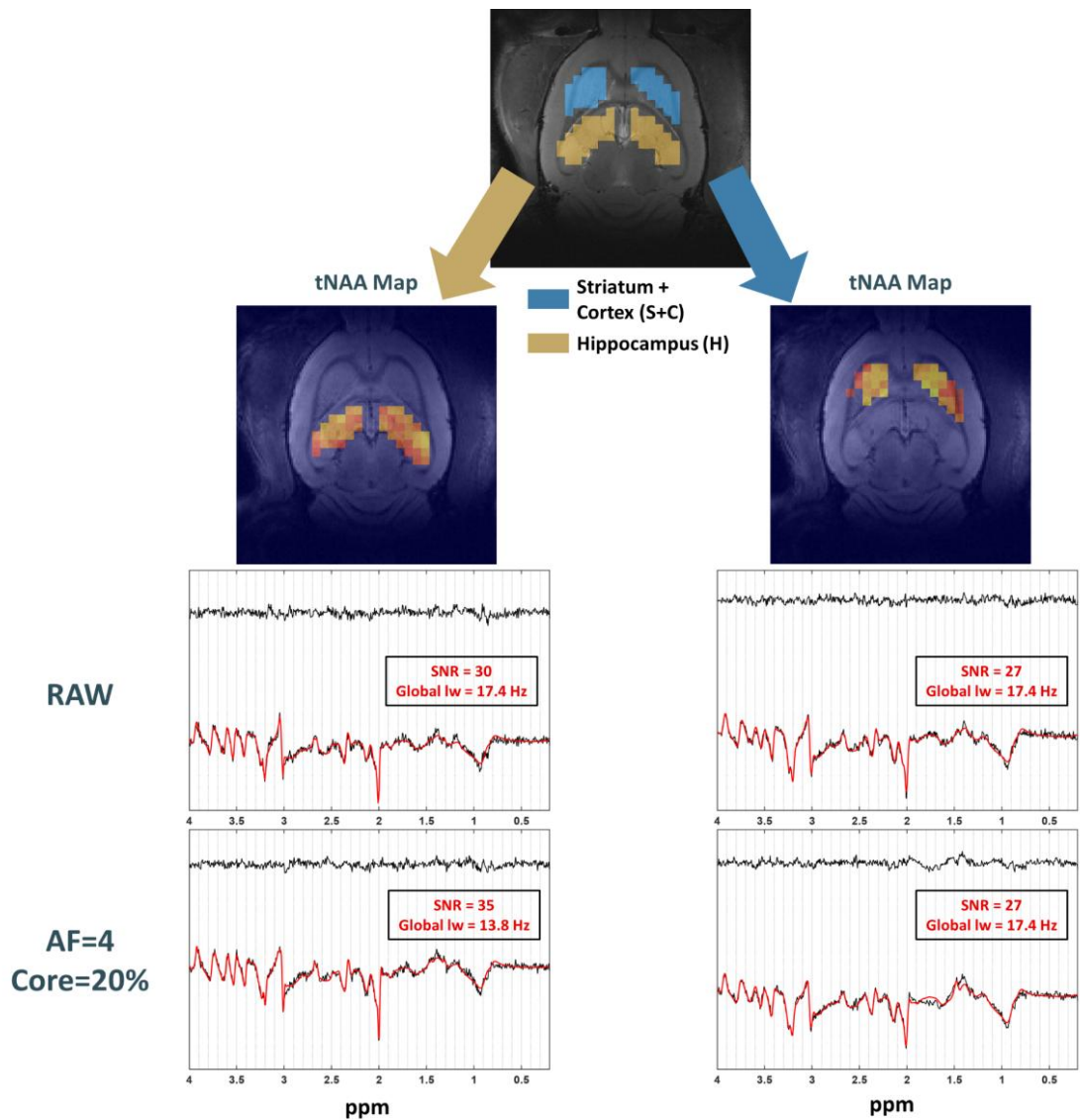

*Figure 2 Selected brain segmentation for the hippocampus and a mix of striatum and cortex. The segmentation was applied on a tNAA metabolite map with a representative spectrum in each region for illustration purposes. Both spectra were obtained at 14.1T with 1 average, with the RAW configuration and with CS, with AF = 4 and Core = 20%.*

# 3. Results

## 3.1. CS Optimization

**Pre-Reconstruction PSF Study and Analysis**

To evaluate the effect of CS on the spatial resolution of the MRSI datasets, the PSF was generated from the *k*-space sampling of each CS configuration (AF and *Core* tests in Figure 3 and 4, respectively;

Supplementary Figure 3 combined). Comparison with the RAW PSF showed an increasing trend for the maximum lobe amplitude and the FWHM when the AF increases (Figure 3, Supplementary Figure 3). At the highest AF tested, the lobes reached an amplitude of up to 12% of the PSF maximum value, and the width was equal to 2.7 times the voxel size, resulting in an effective voxel size of $2.1 \times 2.1$ mm as opposed to $1.39 \times 1.39$ mm for the RAW. The behavior of the lobe amplitude was not linear with the *Core* size: the lobes had a larger reach (up to 6 times the voxel size) for configurations with *Core* value equal and smaller to 10% (Figure 7). This seemed like an outlier case caused by conflicting reconstruction variables such as the regularization. An increase of the FWHM was noted with the *Core* size (Figure 4), but the increasing trend was less pronounced compared with the AF (Figure 4).

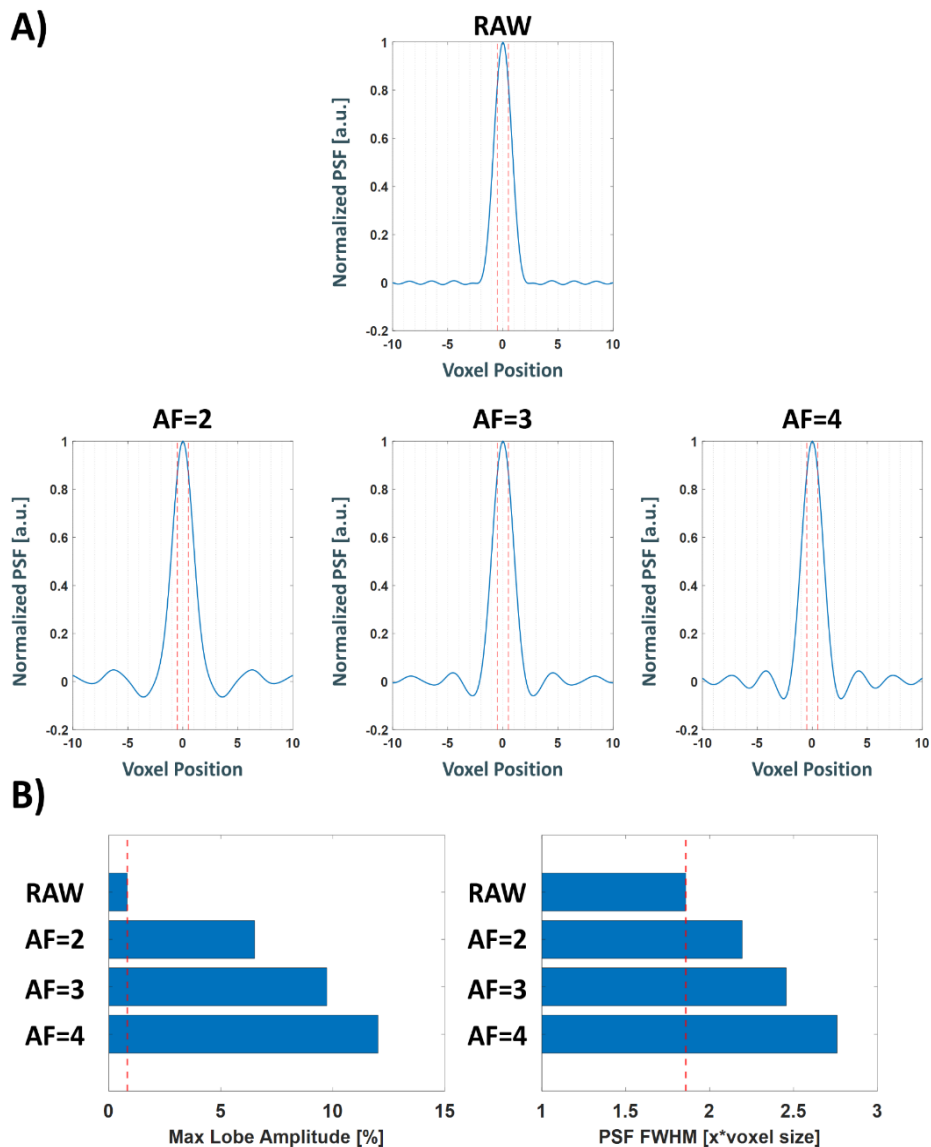

*Figure 3 A) Normalized PSF of the different AF configurations (Core fixed at 20%). The PSF is plotted as a function of the relative matrix position with respect to the signal emitting voxel (i.e. x = 1 means one voxel away from the voxel of origin). The red dotted limitations represent the signal emitting voxel. The application of higher AF tend to increase the number and intensity of the lobes B) A barplot of the maximum lobe amplitude and the FWHM of the PSF of each configuration. Amplitudes are expressed in percentage of the maximum PSF value and the FWHM is expressed in voxel magnification factor (effective size = FWHM*voxel size). The dotted red line serves as reference from the PSF FWHM of the RAW acquisition. The increase of FWHM and lobe amplitude follows the increase of AF.*

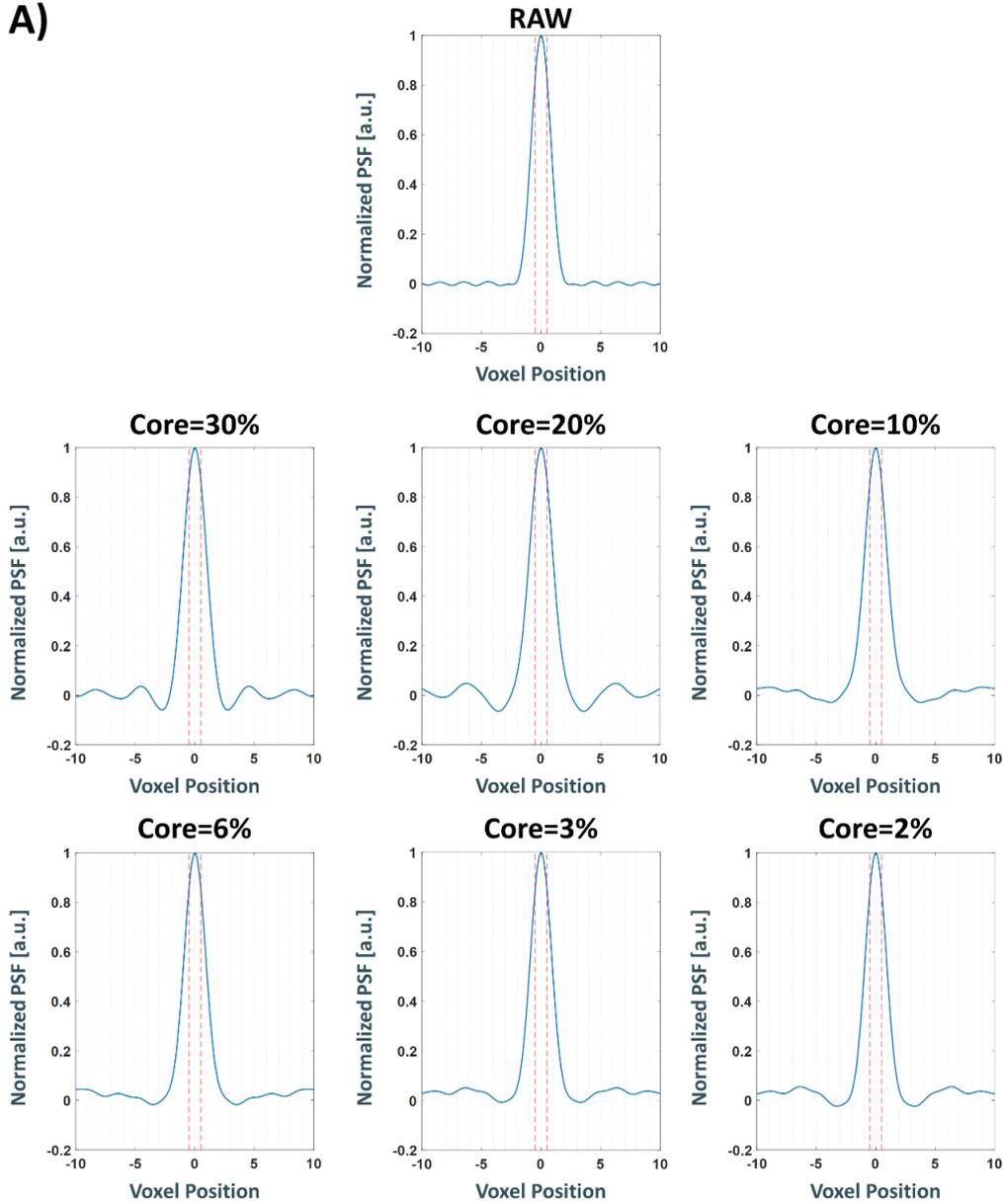

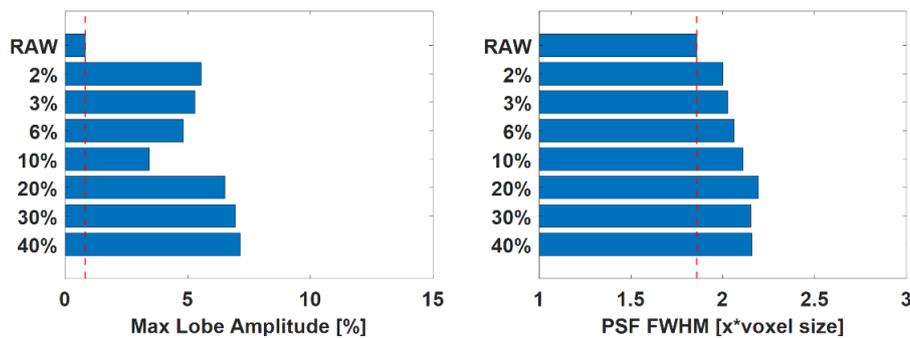

*Figure 4 Normalized PSF of the different Core configurations (AF fixed at 2). A) The PSF is plotted as a function of the position with respect to the signal emitting voxel (i.e. x = 1 means one voxel away from the voxel of origin). The red dotted limitations represent the signal emitting voxel. The lobes amplitude seemed to be lower as the Core is reduced, but the spread is larger (2 voxels to 6 voxels long) B) A barplot of the maximum lobe amplitude and the FWHM of the PSF of each configuration. Amplitudes are expressed in percentage of the maximum PSF value and the FWHM is expressed in voxel magnification factor (effective size = FWHM*voxel size). The dotted red line serves as reference from the PSF FWHM of the RAW acquisition. The lobe amplitude decreases as Core decreases while the FWHM remains similar. The case of 10% seems to be an outlier due to the reconstruction parameter combination, as the spread of the lobe remains long compared to other configurations.*

**QC Analysis**

The mean $SNR_t$ and the mean linewidth for each AF and *Core* size is illustrated in Figure 5. A power relation was fitted on the boxplots to describe the behavior of the $SNR_t$ with respect to the AF ($y = a \cdot AF^x$). A value of x = 0.47 was estimated, suggesting that the gains in $SNR_t$ are due to the acquisition time reduction rather than an effect of CS on the SNR. For the *Core* size, the changes in the mean $SNR_t$ were negligible. The linewidth was not affected by any of the two parameters of the CS, which was expected due to the independence of the *k*-space sampling and the spectral resolution of the MRSI acquisition.

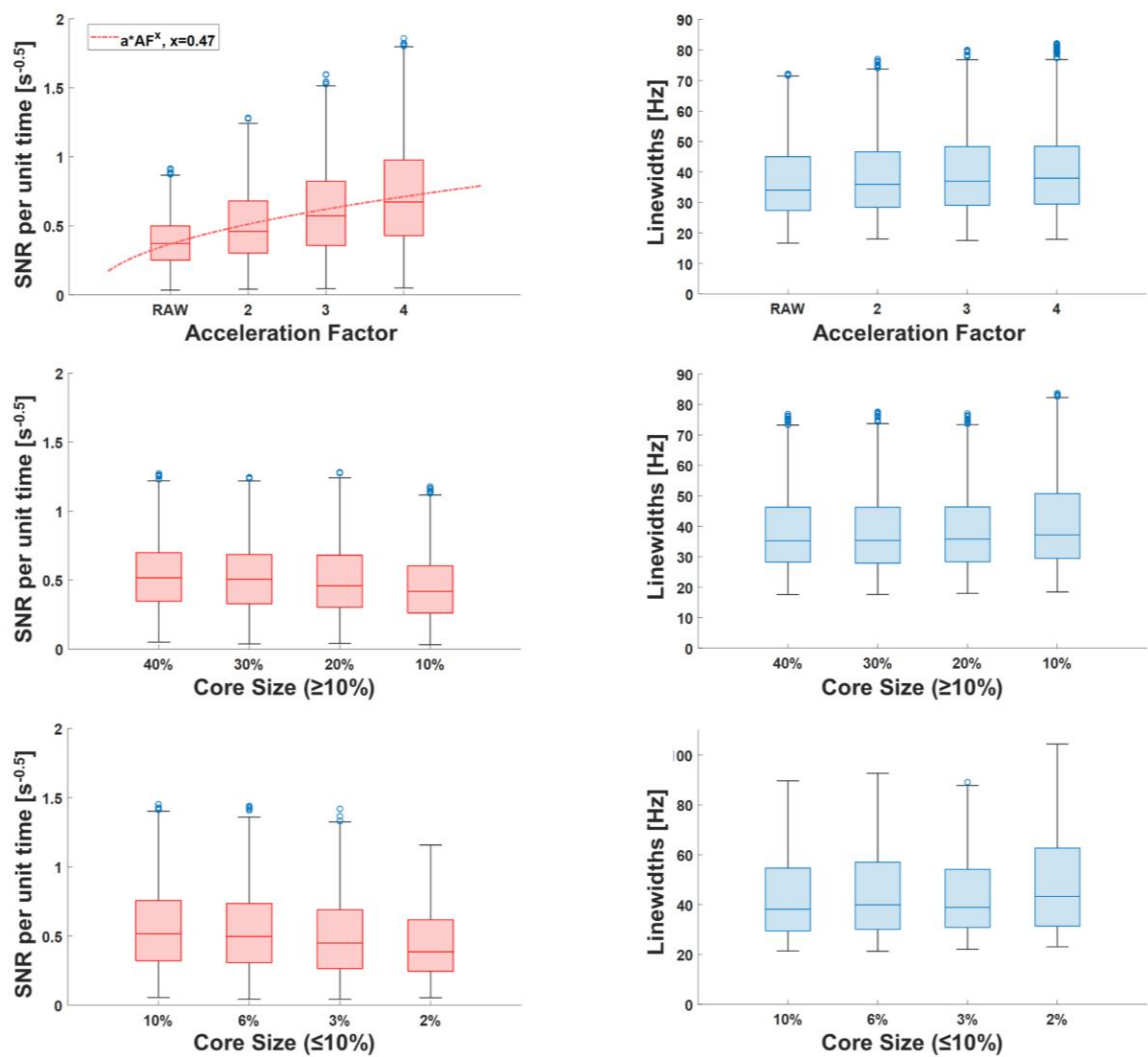

*Figure 5 Box plots of the SNR per unit of time (left column) and the linewidths (right column) as a function of each configuration (from top to bottom: changes in AF, changes in Core size above 10% and below 10% / outliers in blue data points). For the plot of the different AFs, a power law regression (y = a*AFx) is fitted on the mean SNR per unit time value. As the fit follows closely a square root law (x = 0.47), the gains in SNR per unit of time are principally due to the reduced acquisition time. No significant difference was noted in average linewidth with each configuration.*

**Lipid contamination with CS in the rat brain**

As a proof of concept, an acquisition with AF = 4 and *Core* = 6% was conducted and compared with an acquisition with AF = 4 and *Core* = 20%, in order to evaluate, within preclinical acquisition, the lipid contamination due to high CS accelerations induced spatial aliasing reported in clinical MRSI[31]. The results (Figure 6) showed that the spectra acquired with AF = 4 and *Core* = 6% were heavily contaminated, to the point where LCModel struggled with the fitting, as opposed to AF = 4 and *Core* = 20% where the spectral quality was similar to what was reported in the other configurations. This translated into a more sparse coverage that rendered the regional concentration estimates study more complicated and less reliable. Therefore, while AF > 2 can be used with a small cost of coverage and precision, the results show that the *Core* size should be relatively large (between 10 and 20% for 2D acquisition) to avoid unexpected lipid contamination.

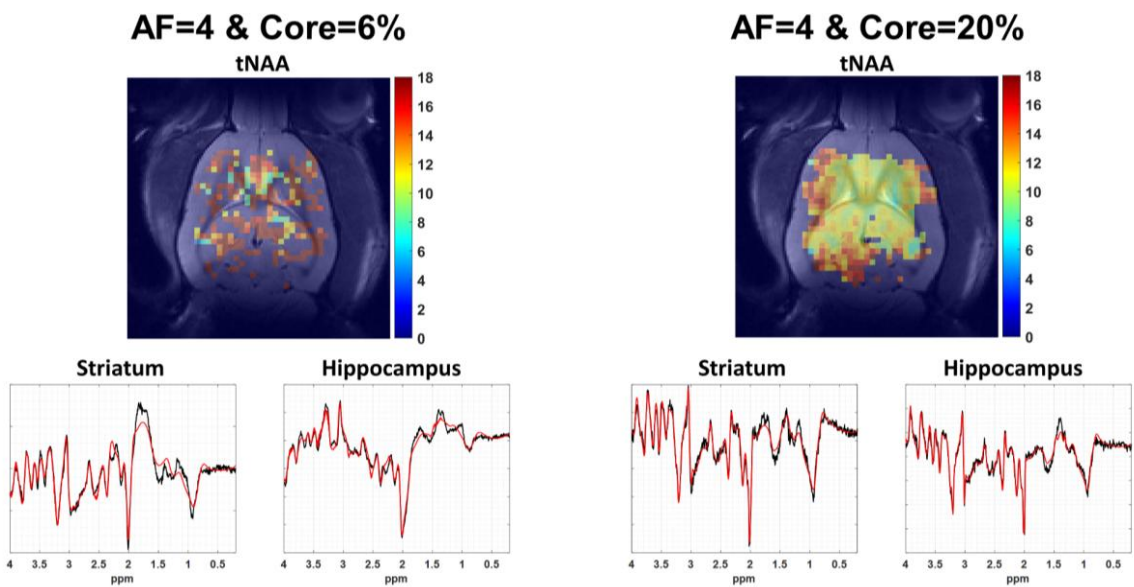

*Figure 6 Metabolic map of tNAA and representative spectrum in the striatum and hippocampus, for AF = 4 and Core = 6% and AF = 4 and Core = 20%. The coverage is significantly reduced for AF = 4 and Core = 6% and the spectra showed heavy contamination around 2 and 1 ppm that impacted the metabolite signal.*

**Spectral Fitting, Metabolite Maps and Estimates**

Each configuration (AF and *Core* size) provided reliable spectra that were successfully fitted by LCModel (Figures 7-9). While no significant visual differences between 4 and 2 ppm were noted between the RAW and the different AF and *Core* parameters, the RAW spectra and fit residuals between 2 and 0 ppm were altered when using CS (in general at *Core* size 10% and below combined with high AF), mainly due to increased lipid contamination. The metabolic maps of tNAA, tCho, and Ins are displayed for all different setups in Figures 10-12. The maps showed a more granular and less smooth distribution when CS was applied especially at low *Core* size, as well as some differences in coverage due to the noise-like aliasing caused by the increase in sampling randomness[31]. A mean overall coverage loss of 10% was observed with the different AF, while the loss was more gradual with the *Core* size (3% for *Core* = 40% to 16% for *Core* = 2%). This was especially noticeable for tCho where the metabolite map coverage changed significantly when applying AF = 3 or higher or when the *Core* ≤ 10%. The CRLBs maps of tNAA, tCho and Ins (Supplementary Figures 4-6) support this trend as an increase of the CRLBs for the three metabolites was observed on the borders of the maps. The CRLBs of tCho stand out as more voxels on the central part of the maps reach CRLB values between 30% and 40% when using CS, impacting the overall coverage.

Figure 13 illustrates the mean relative concentration estimates, as well as the mean CRLB, of tNAA, Gln, Glu, Ins, Tau, and tCho at the two regions of interest (for all AFs, *Core* ≥ 10% and *Core* ≤ 10%, respectively). Despite the decrease in coverage, no significant difference on metabolite estimation on the voxels passing the QC filter was noted between the RAW and any configuration, with the highest coefficient of variation computed being 5% for AFs (for tCho), 4% for *Core* ≥ 10% (for tCho and Tau) and 8% for *Core* ≤ 10% (for tNAA). However, a consistent increase of the SD was found when CS was used, regardless of the parameters chosen. This was coherent with the visual impression of granularity found in the metabolic maps. No particular trend was observed with the AF, as the increase of SD was 50% on average for all metabolites regardless of what AF was applied. The SD increased consistently as the *Core* size was reduced, capping at around a 100% increase. It is worth noting that this increase was calculated with respect to the RAW data of each test. As such, due to the difference in rats used for each *Core* test, the value of the SD may differ between the RAW of each test. The CRLBs did not differ significantly across the regions of interest (hippocampus and striatum).

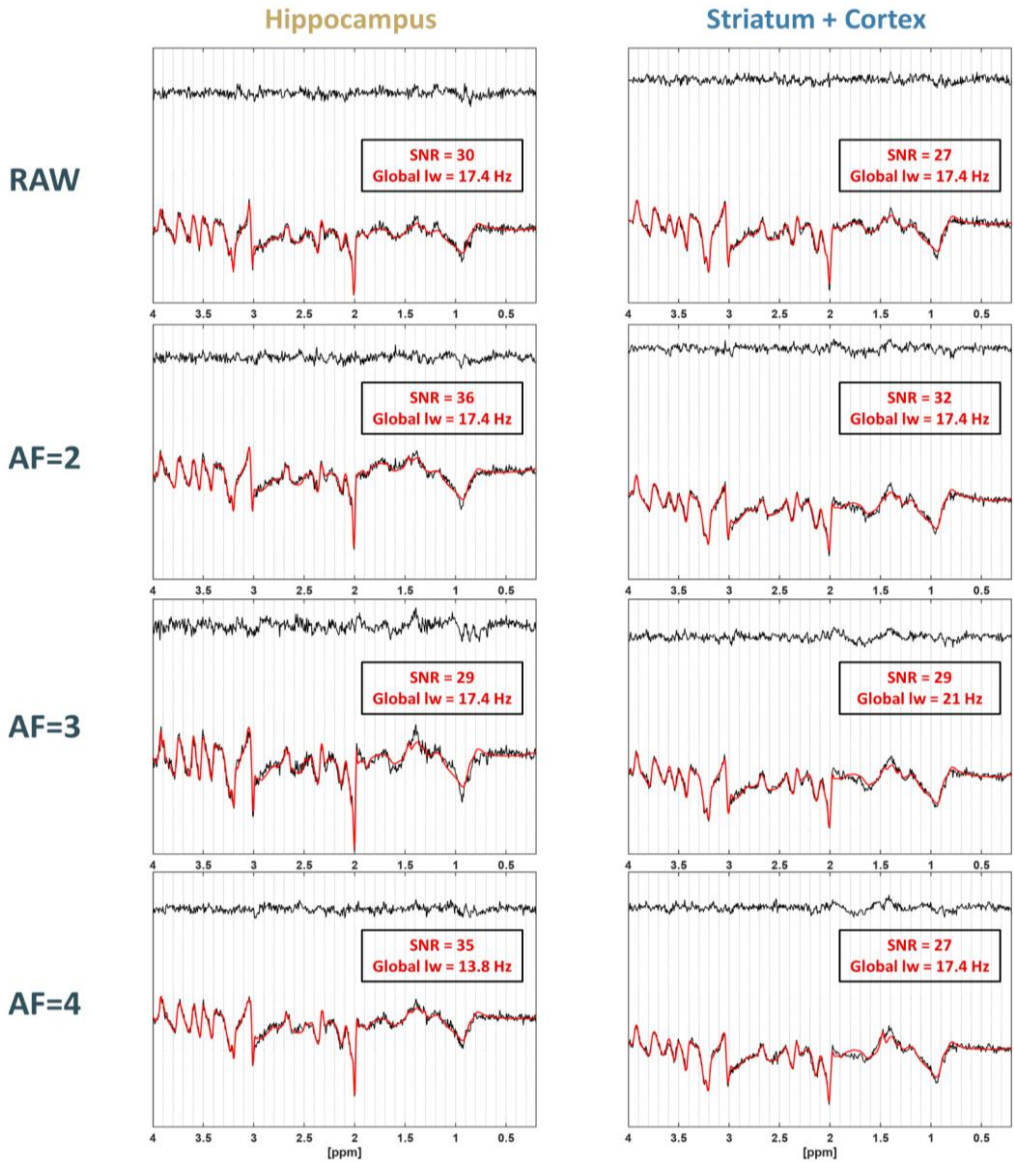

*Figure 7 Representative in vivo spectra acquired at 14.1T, in both hippocampus (left column) and striatum (right column) for the different AF configurations (Core size fixed at 20%). The SNR and linewidth estimated by LCModel are displayed in the box. The residuals, in black above the fit, show the appearance of patterns in the region below 2ppm, while no significant alteration in the residual profile is seen above 2ppm.*

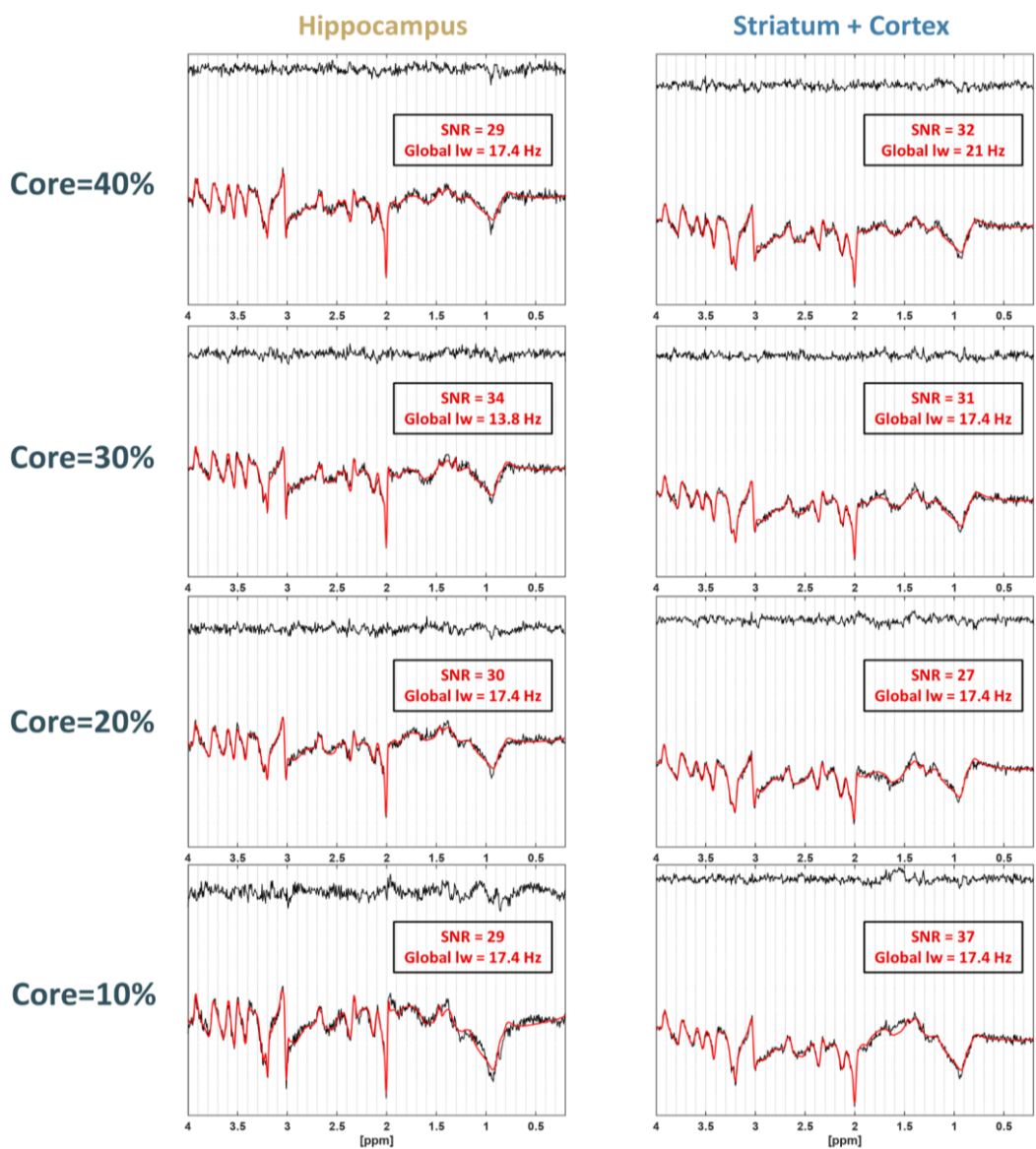

*Figure 8 Representative in vivo spectra acquired at 14.1T, in both hippocampus (left column) and striatum (right column) for the different Core configurations (≥10%; fixed AF at 2). The SNR and linewidth estimated by LCModel are displayed in the box. The residuals, in black above the fit, show the appearance of patterns in the region below 2ppm, while no significant alteration in the residual profile is seen above 2ppm.*

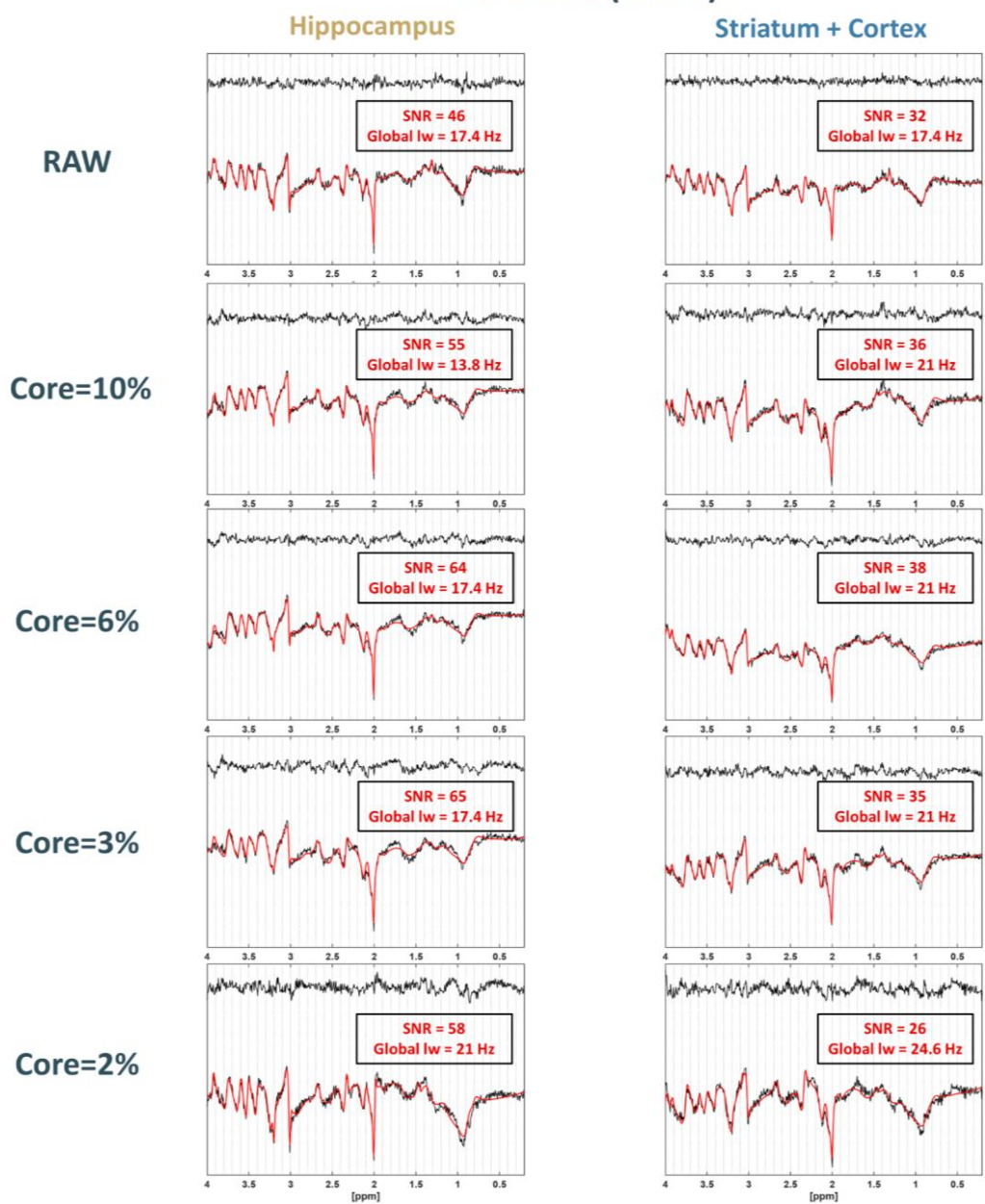

*Figure 9 Representative in vivo spectra acquired at 14.1T, in both hippocampus (left column) and striatum (right column) for the different Core configurations (≤10%). The SNR and linewidth estimated by LCModel are displayed in the box. The residuals, in black above the fit, show the appearance of patterns in the region below 2ppm, while no significant alteration in the residual profile is seen above 2ppm.*

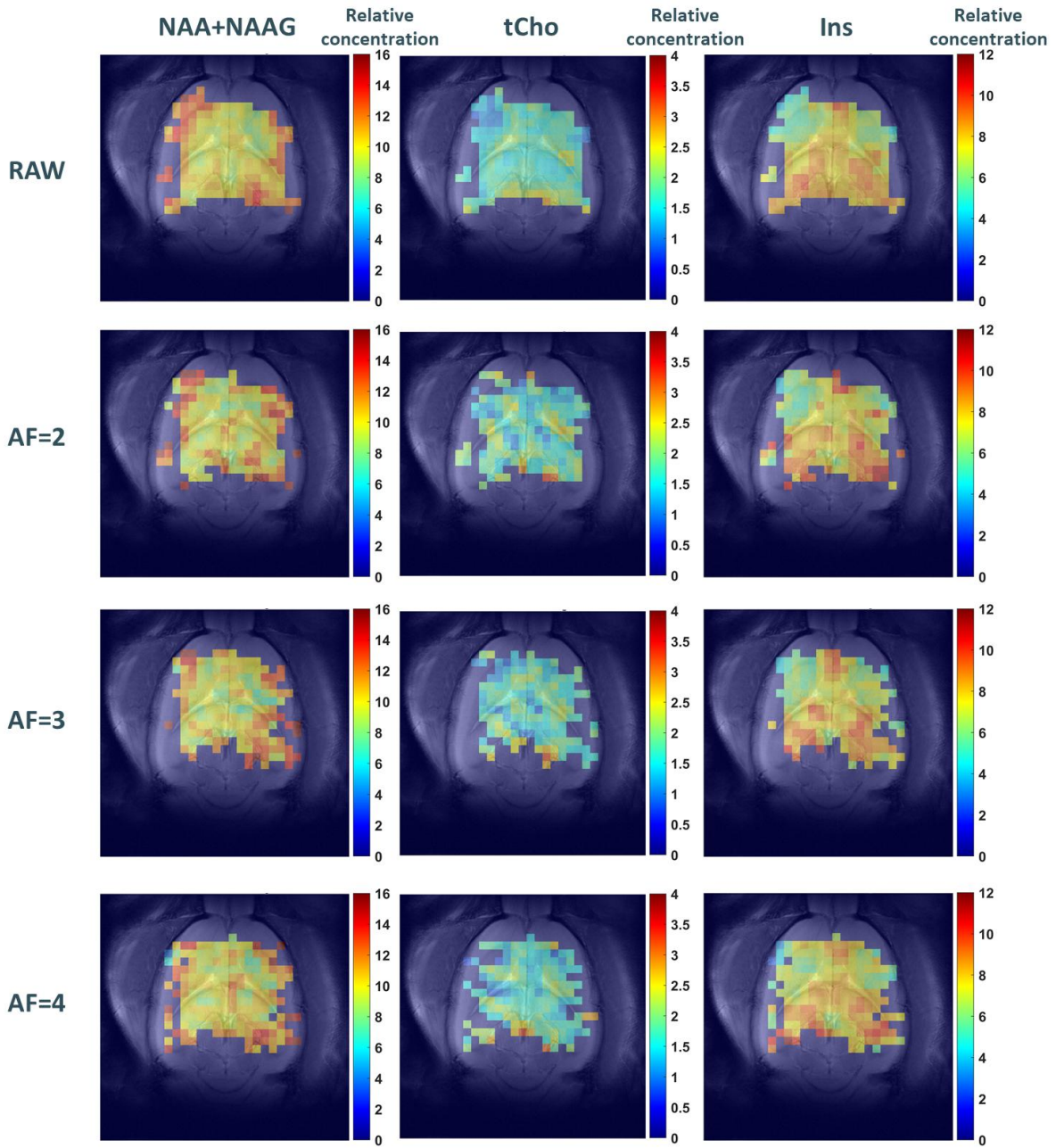

*Figure 10 Representative in vivo metabolite maps acquired at 14.1T of tNAA, tCho and Ins, for the RAW and different AFs configuration (Core Size fixed at 20%). The metabolite maps were superimposed to the corresponding anatomical image with the MRS4Brain Toolbox. An alteration of the coverage and the appearance of a granular structure on the metabolic maps can be seen with each AF compared to RAW. The scales correspond to LCModel outputs when referenced to tCr by setting its concentration to 8 mmol/kgww.*

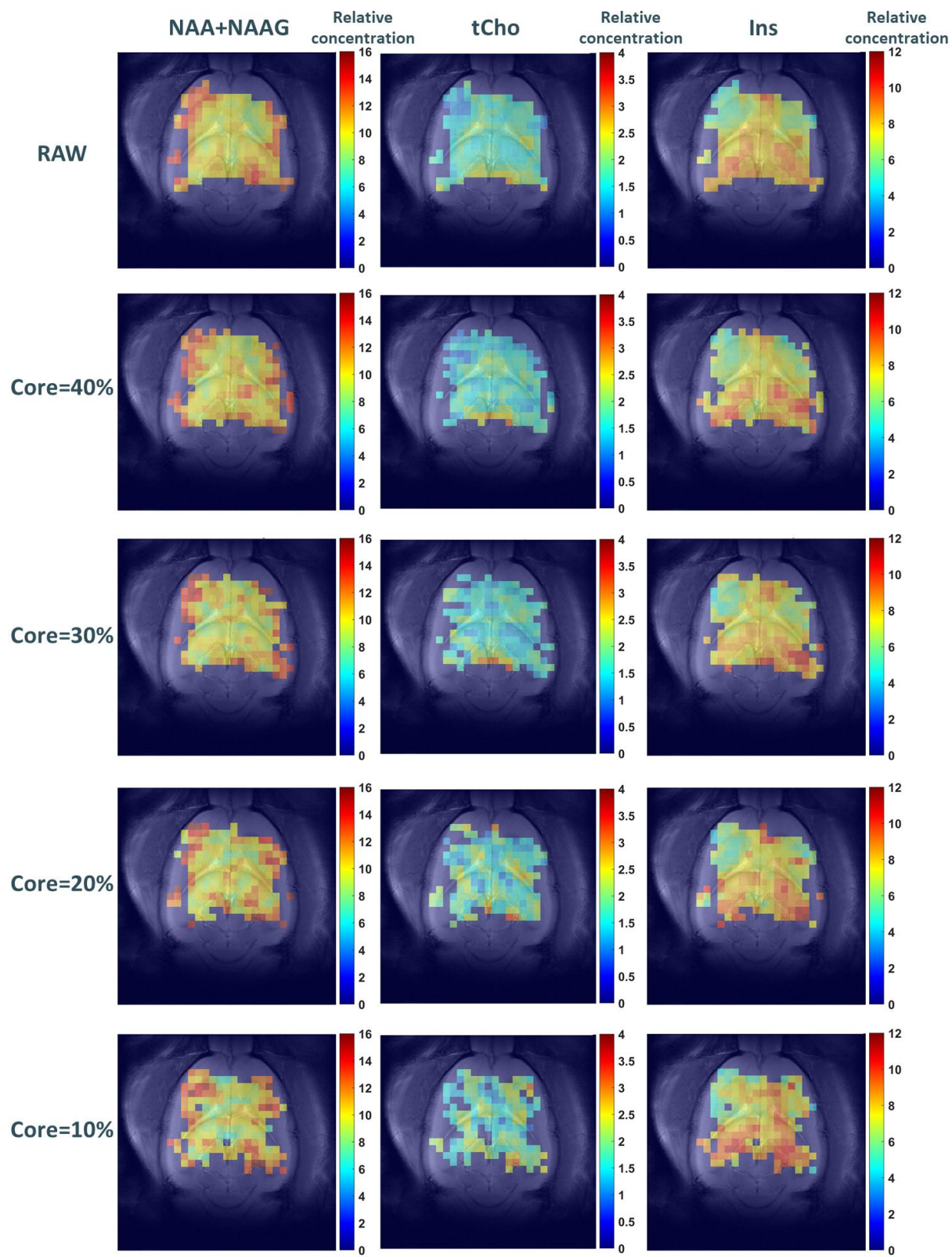

*Figure 11 Representative in vivo metabolite maps acquired at 14.1T of tNAA, tCho and Ins, for the RAW and different Core configuration (≥10%; AF fixed at 2). The metabolite maps were superimposed to the corresponding anatomical image with the MRS4Brain Toolbox. An alteration of the coverage and the appearance of a granular structure on the metabolic maps can be seen with each Core size compared to RAW. The scales correspond to LCModel outputs when referenced to tCr by setting its concentration to 8 mmol/kgww.*

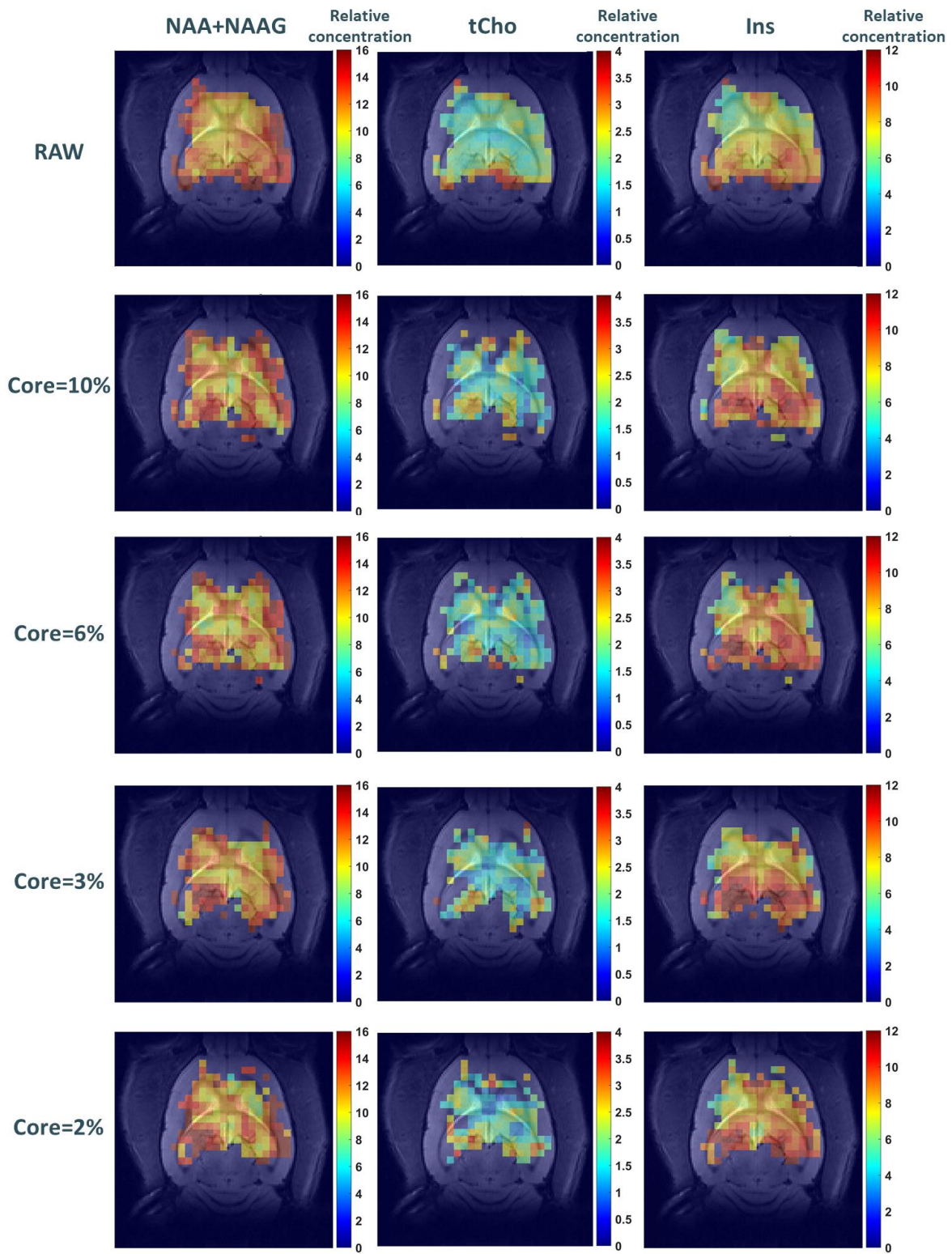

*Figure 12 Representative in vivo metabolite maps acquired at 14.1T of tNAA, tCho, and Ins, for the RAW and different Core configuration (≤10%; AF fixed at 2). An alteration of the coverage and the appearance of a granular structure on the metabolic maps can be seen with each Core size compared to RAW. The metabolite maps were superimposed to the corresponding anatomical image with the MRS4Brain Toolbox. The scales correspond to LCModel outputs when referenced to tCr by setting its concentration to 8 mmol/kgww.*

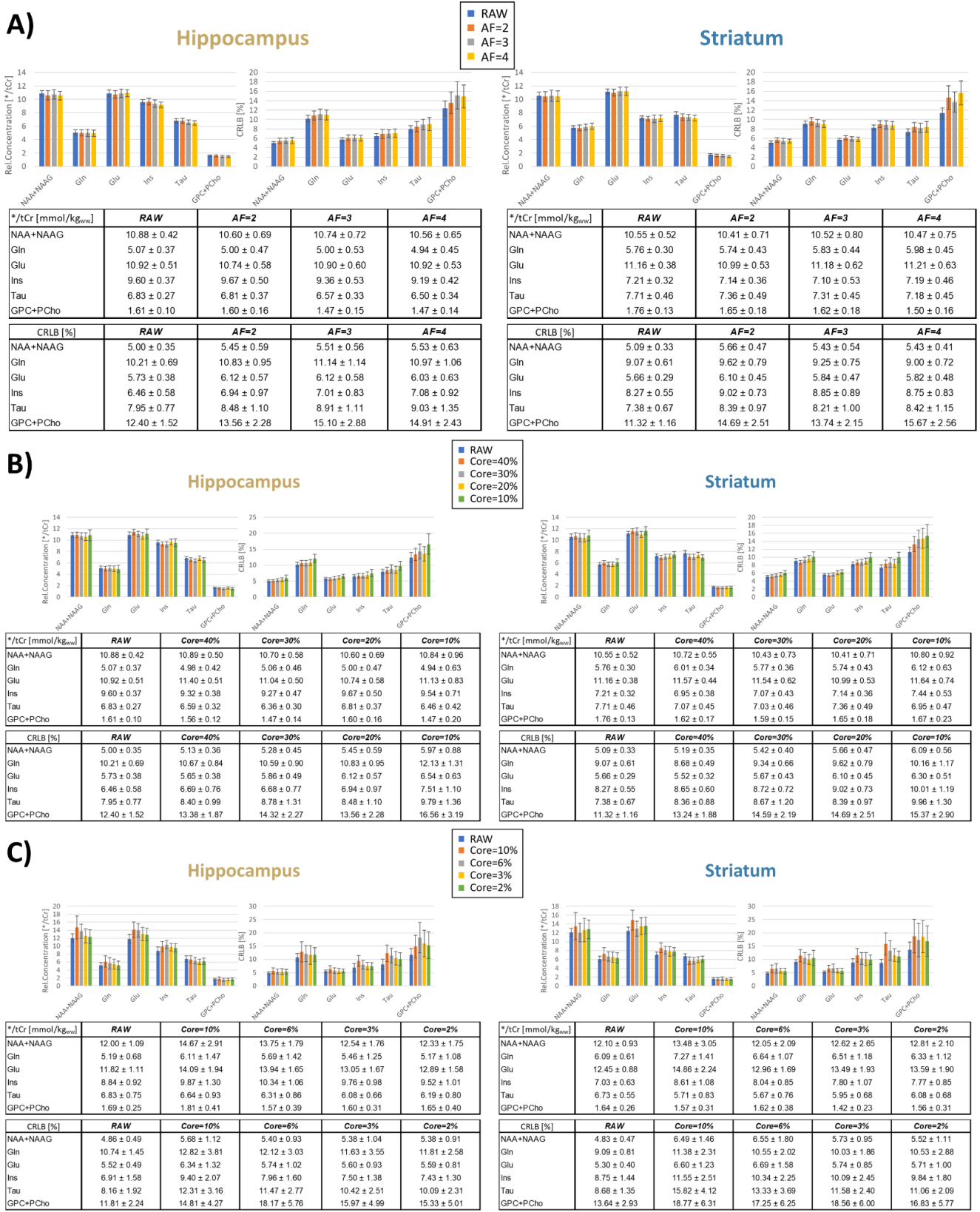

*Figure 13 Quantitative and CRLB assessment in each brain region (at 14.1T, mean of the mean over 6 rats, per average), for A) each AFs configuration (Core Size fixed at 20%), B) Core Size ≥10% configuration and C) Core Size ≤10% configuration (AF fixed at 2 for both B and C). No significant difference of the estimates (tables of the 1st row) were found. The SDs are higher on the Core <10% study table as the rats used for the averaging were different compared to the two other cases*

.

## 3.2. High-Resolution FID-MRSI with CS

**QC Maps and Metrics**

The SNR maps and normalized SNR distribution (Figure 14A) further underline the preservation of the spectra quality as the distribution remains unchanged between the resolutions. An increase of the non-normalized SNR can be observed in the maps of $47 \times 47$ due to the higher averaging managing to compensate for the loss of signal caused by the smaller voxel size (Supplementary Table 2). The effective voxel size for each configuration was also computed taking into account the PSF. The CRLB maps are used in the *MRS4Brain Toolbox* as QC. Figure 14B highlights the robustness of the acquisitions via the CRLB maps, as very few visual differences between each configuration were detected.

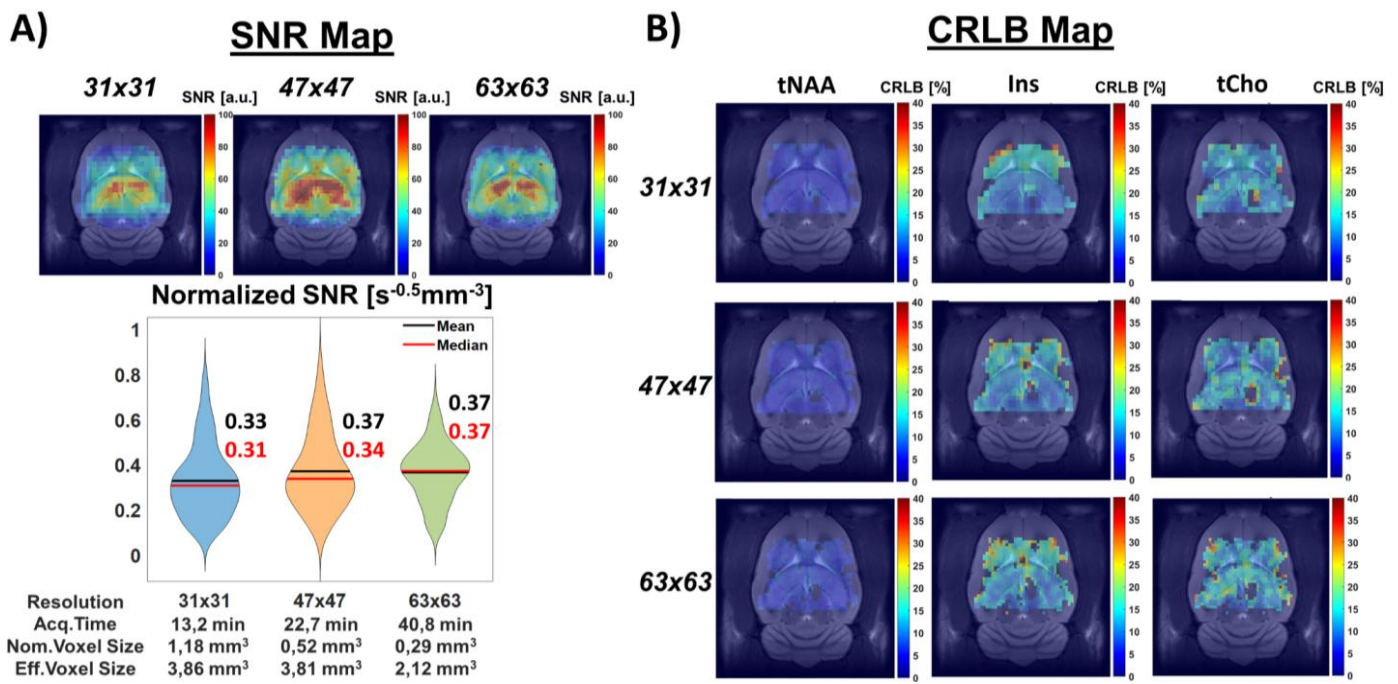

*Figure 14 A) SNR maps and normalized SNR distribution for the 31x31, 47x47, and 63x63 configurations. The SNR map 47x47 displays a higher SNR due to the averaging (31x31: 1avg/ 47x47 and 63x63: 3avg) compensating the decrease in nominal voxel size. The relevant parameters for the normalization are displayed below the violin plot, while the mean and median value of the distributions are reported in black and red, respectively. The distribution of the normalized SNR show no significant alteration between configurations. B) Representative CRLB maps of tNAA, tCho, and Ins, for the 31x31, 47x47 and 63x63 configurations, acquired at 9.4T. The metabolite maps were superimposed to the corresponding anatomical image with the MRS4Brain Toolbox. No significant difference was noted in the distribution between the configurations. The scales correspond to a CRLB limit set at 40%.*

**Spectral Fitting, Metabolite Maps and Estimates**

To evaluate the potential of CS with the chosen optimized parameters (AF = 4; *Core* = 20%) for an increase of in-plane resolution, the metabolic maps of tNAA, Ins, and tCho were acquired with each resolution configuration (Figure 15A). The metabolic distribution remained relatively unchanged qualitatively: the Ins regional difference between hippocampus and striatum was preserved between the configurations. It is worth pointing out that, for the $63 \times 63$ acquisition, some voxels in the middle of the maps were filtered out by the QC due to a slight decrease in SNR as the decrease of the voxel size does not compensate for the averaging. The spectra in each brain region (Figure 15B) were reliably fitted regardless of the resolution, with minimal difference between the residuals.

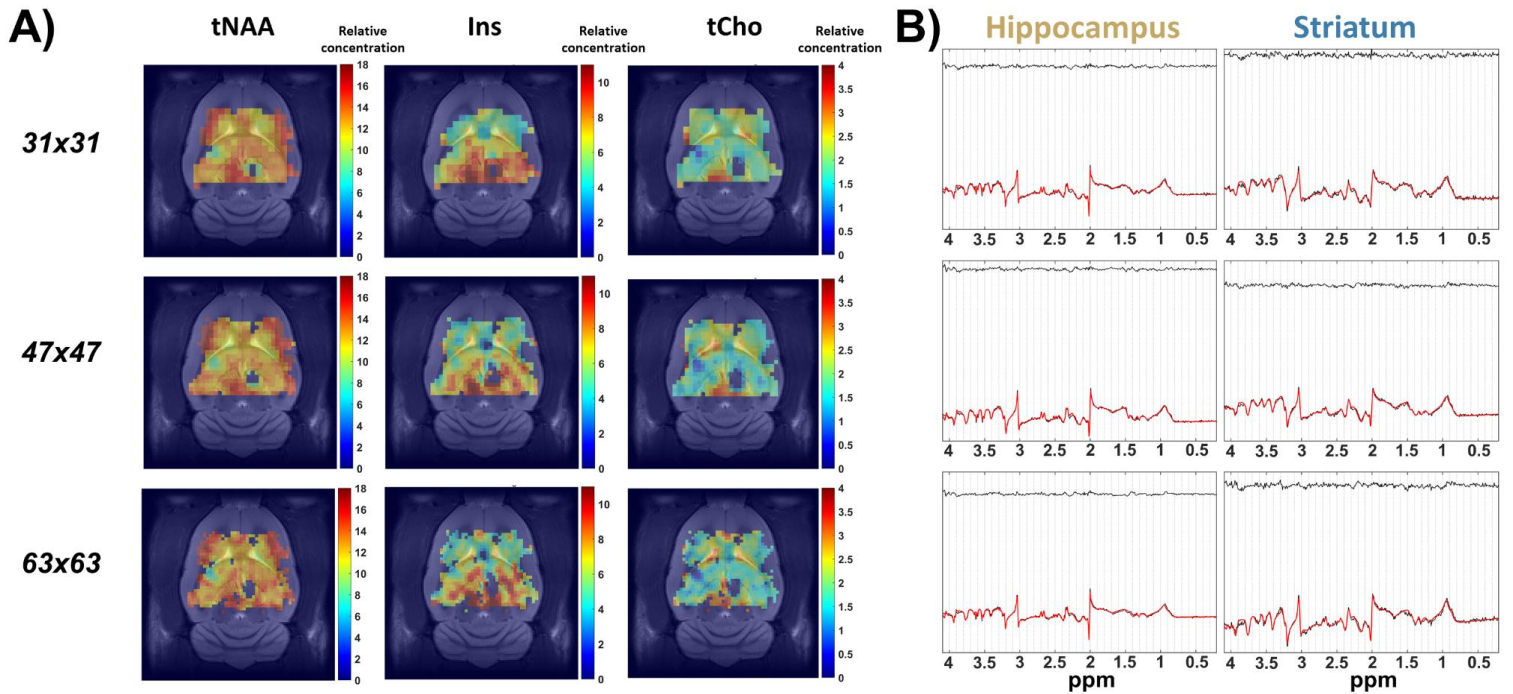

*Figure 15 A) Representative in vivo metabolite maps of tNAA, tCho and Ins, for 31x31, 47x47 and 63x63 configuration, acquired at 9.4T. The metabolite maps were superimposed to the corresponding anatomical image with the MRS4Brain Toolbox. The coverage was slightly altered in the central part of the maps (especially in the 63x63) due to a reduction in SNR. The metabolic maps retained their concentration distribution and profile throughout the configurations. The scales correspond to LCModel outputs when referenced to tCr by setting its concentration to 8 mmol/kgww. B) Representative spectra in both hippocampus (left column) and striatum (right column) for 31x31, 47x47 and 63x63 configuration. No alteration was noted in both the fit and the residuals.*

The mean relative concentration estimates for tNAA, Gln, Glu, Ins, Tau, and tCho (Figure 16) were similar between the three resolutions, with the highest variation being 5% for tCho in the hippocampus and 5% for the Ins in the striatum. The CRLBs were also not significantly altered, echoing the visual impression from the metabolic and quality maps. The higher SNR with the $47 \times 47$ configuration translated into a lower mean CRLB (not statistically significant) for the metabolites of interest, especially in the striatum region.

# Hippocampus

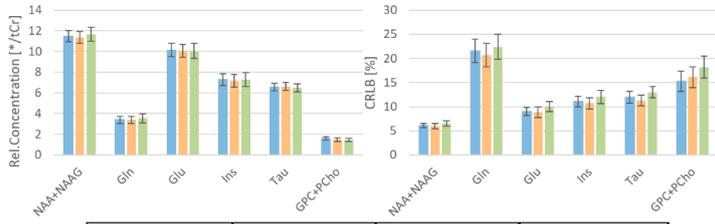

| */tCr [mmol/kg_ww] | 31x31 | 47x47 | 63x63 |
|---|---|---|---|
| NAA+NAAG | 11.48 ± 0.55 | 11.37 ± 0.56 | 11.67 ± 0.68 |
| Gln | 3.38 ± 0.33 | 3.39 ± 0.34 | 3.55 ± 0.43 |
| Glu | 10.14 ± 0.63 | 10.05 ± 0.62 | 10.07 ± 0.71 |
| Ins | 7.27 ± 0.59 | 7.18 ± 0.61 | 7.26 ± 0.66 |
| Tau | 6.55 ± 0.37 | 6.61 ± 0.38 | 6.48 ± 0.41 |
| GPC+PCho | 1.59 ± 0.15 | 1.48 ± 0.18 | 1.45 ± 0.17 |

| CRLB [%] | 31x31 | 47x47 | 63x63 |
|---|---|---|---|
| NAA+NAAG | 6.09 ± 0.49 | 5.96 ± 0.53 | 6.58 ± 0.55 |
| Gln | 21.58 ± 2.40 | 20.72 ± 2.42 | 22.38 ± 2.62 |
| Glu | 9.00 ± 0.81 | 8.83 ± 1.11 | 10.00 ± 1.07 |
| Ins | 11.04 ± 1.10 | 10.69 ± 1.16 | 12.05 ± 1.38 |
| Tau | 11.95 ± 1.17 | 11.28 ± 1.08 | 12.99 ± 1.19 |
| GPC+PCho | 15.26 ± 2.13 | 16.12 ± 2.14 | 18.19 ± 2.31 |

# Striatum

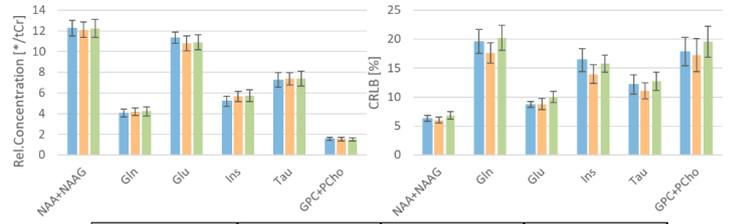

| */tCr [mmol/kg_ww] | 31x31 | 47x47 | 63x63 |
|---|---|---|---|
| NAA+NAAG | 12.25 ± 0.75 | 12.09 ± 0.75 | 12.21 ± 0.88 |
| Gln | 4.05 ± 0.39 | 4.17 ± 0.36 | 4.22 ± 0.43 |
| Glu | 11.33 ± 0.55 | 10.81 ± 0.72 | 10.90 ± 0.71 |
| Ins | 5.19 ± 0.50 | 5.66 ± 0.48 | 5.74 ± 0.58 |
| Tau | 7.25 ± 0.71 | 7.37 ± 0.60 | 7.40 ± 0.73 |
| GPC+PCho | 1.56 ± 0.16 | 1.54 ± 0.18 | 1.49 ± 0.17 |

| CRLB [%] | 31x31 | 47x47 | 63x63 |
|---|---|---|---|
| NAA+NAAG | 6.30 ± 0.50 | 6.02 ± 0.48 | 6.83 ± 0.62 |
| Gln | 19.59 ± 2.06 | 17.62 ± 1.76 | 20.21 ± 2.16 |
| Glu | 8.70 ± 0.50 | 8.80 ± 0.94 | 9.98 ± 0.94 |
| Ins | 16.39 ± 1.98 | 13.95 ± 1.61 | 15.74 ± 1.49 |
| Tau | 12.17 ± 1.69 | 11.08 ± 1.37 | 12.74 ± 1.57 |
| GPC+PCho | 17.81 ± 2.46 | 17.23 ± 2.87 | 19.53 ± 2.67 |

*Figure 16 Quantitative and CRLB assessment in each brain region, for 31x31, 47x47 and 63x63 resolution (mean of the mean over 5 rats, per average). No significant variation of the estimates (tables of the 1st row) were found, while a slight decrease of the CRLB in 47x47 (caused by a higher SNR, see Figure 14) compared to 31x31 and 63x63 was noted.*

## 3.3. 3D and Multislice FID-MRSI with CS

**Pre-reconstruction PSF Comparison**

The difference in *k*-space sampling between 3D and Multislice has a noticeable impact on the PSF of each acquisition mode (Figure 17). The lower *Core* value in 3D caused by the limited number of phase encoding steps in the coronal direction (z-axis) lead to an in-plane PSF (x and y axis) with lobes that spreads to longer distances but with a lower amplitude (maximum lobe amplitude was 2.5%) as opposed to higher *Core*. On the other hand, the Multislice acquisition with its higher *Core* value offered a more intense amplitude (maximum at 12%) but with shorter lobes. The FWHM of the PSF of the low *Core* 3D configuration was lower than the high *Core* Multislice (2.3 and 2.8 voxels, respectively), a result in agreement with the optimization done in Section 3.1. The main distinction was made in the coronal direction where, for 3D, a phase encoding process was done, while for Multislice, a slice selection was used. Due to the usage of a slice-selective pulse for the Multislice, the FWHM of the PSF in the z direction is notably smaller than in 3D (1 voxel for Multislice, against 2.4 voxels for 3D). The chemical shift displacement error, however, has to be taken into account for the Multislice: the displacement reported with these spectral parameters was 0.05 mm per part per millions, which corresponds to a maximum shift of 10% of the slice thickness.

## 3D

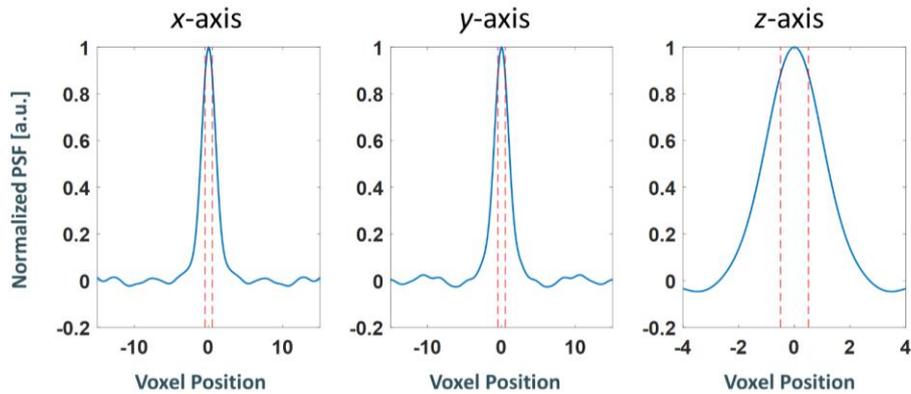

## Multislice

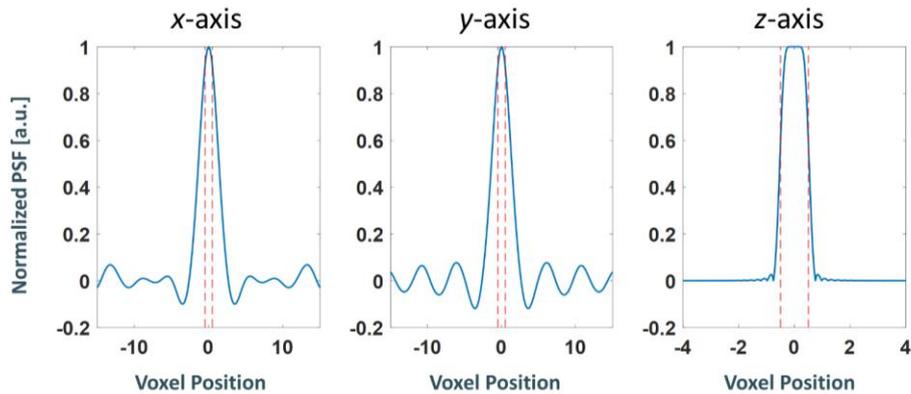

*Figure 17 Projections on the three encoding directions (or two+slice selection for Multislice) of the PSF for the 3D and the Multislice acquisitions. The different CS patterns (shown in Figure 1) have a direct impact on the in-plane (x,y axis) PSF profile. The 3D PSF lob amplitude has a lower intensity but a longer range compared to the Multislice PSF.*

**QC Maps and Metrics**

The SNR distribution along the shared slice (Figure 18A) was relatively similar between 2D and Multislice, but there was an increase of nearly double the mean and median value between 2D and 3D. The difference in in-plane SNR distribution between 2D and Multislice can be explained by the smaller voxel size (in z direction: 2 mm vs 1 mm) in the later modality. The in-plane linewidth distribution (Figure 18B) was more skewed towards higher values for 3D and Multislice. This was expected as the shimming procedure[37] is most effective with smaller volumes of interest. As such, the shimming of the 2D acquisition allowed for an overall smaller linewidth than the adapted procedure for 3D and Multislice.

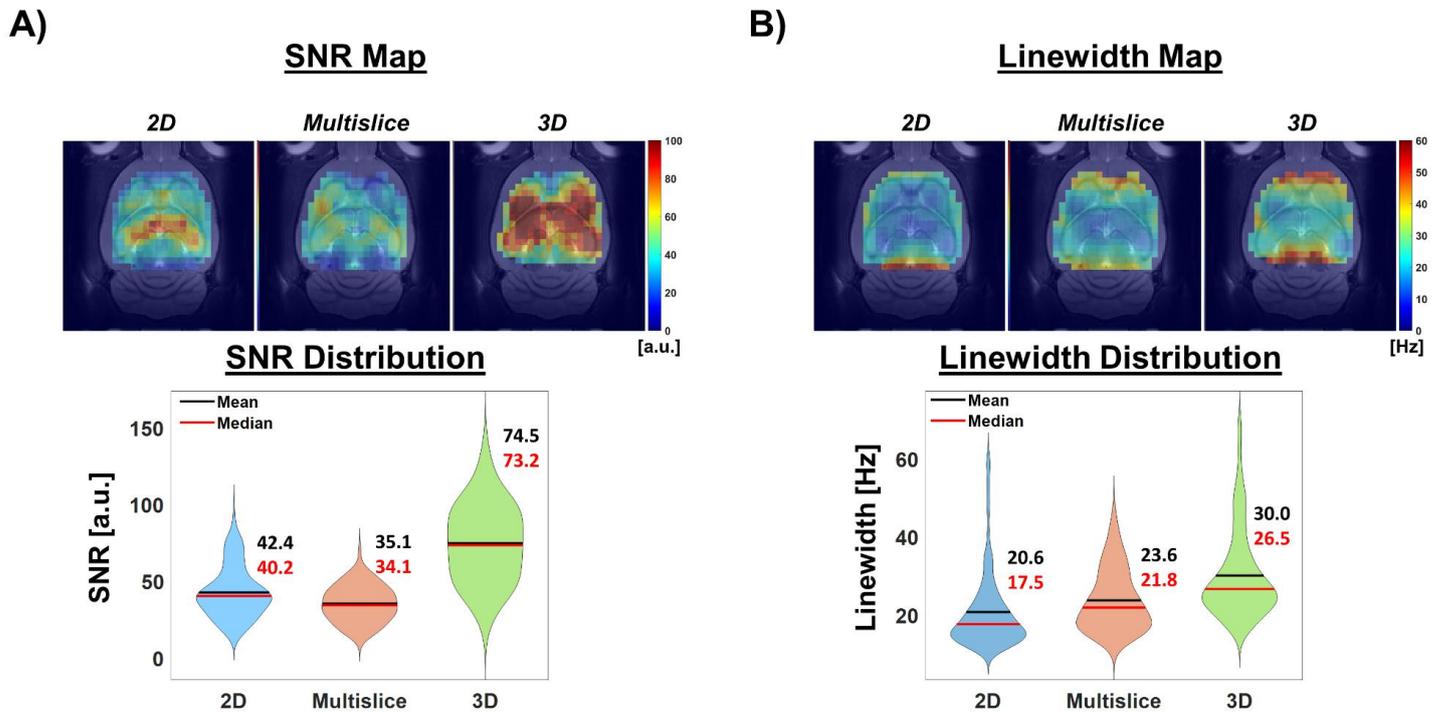

*Figure 18 A) SNR maps and distribution for 2D, Multislice and 3D configuration. The distributions are displayed with the violin plot. For the SNR, no normalization was done as all three methodologies have the same acquisition time. The SNR gains from the 3D modality can be seen through the map and the distribution B) Linewidth maps and distribution for 2D, Multislice and 3D configuration. The distributions are displayed with the violin plot. The Multislice and the 3D are more skewed towards the higher values due to the difference in shimming procedure.*

**Coverage Analysis**

Figure 19 illustrates the Ins distribution as well as a representative spectrum for each step or slice in the coronal direction. Due to the saturation bands configuration and the signal aliasing in 3D caused by the small number of through-plane phase encoding steps, the first two and last two slices were not shown. The two modalities allowed for the acquisition of good-quality spectra in each slice. As expected, the 3D acquired spectra had a higher SNR compared to the Multislice, a decrease in SNR was noted when looking at the slice 6 and 7, corresponding to the ventral part of the rodent brain. This SNR loss translated into a substantial coverage loss in the Ins map for the Multislice technique, going from 65% to 45% on average. The through-slice evolution of the linewidth remained fairly constant, highlighting the efficiency of the shimming procedure for 3D MRSI acquisition.

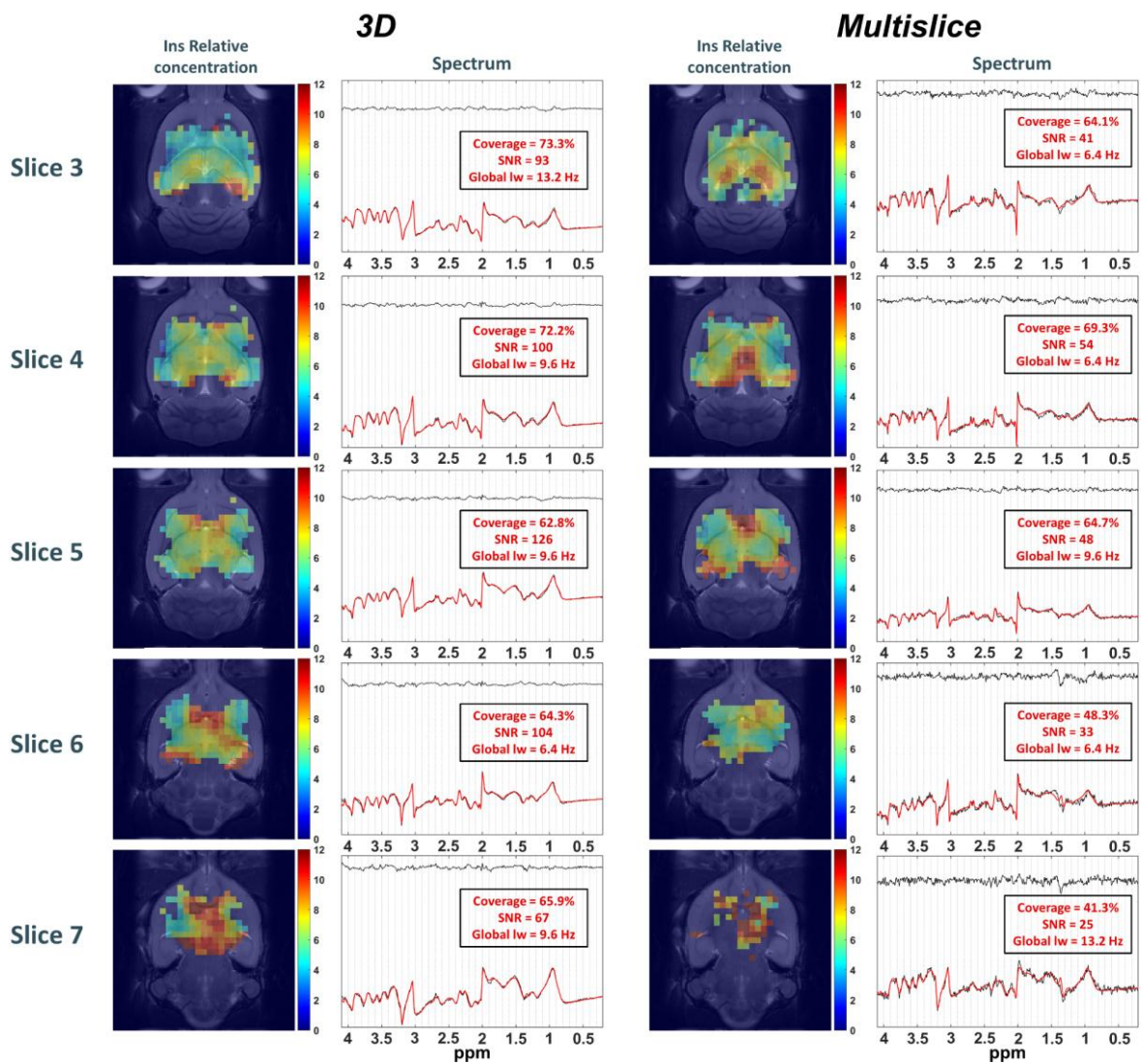

*Figure 19 Ins metabolic map and representative spectrum acquired at 9.4T, with the 3D and Multislice modality on the different slices. The coverage (reported with respect to Ins map), the SNR and global linewidth of the spectra are framed for each slice. A common drop of SNR was noticed between slice 6 and 7 for both modality, while a significant reduction of coverage was observed with the Multislice on these slices. The metabolite maps were superimposed to the corresponding anatomical image with the MRS4Brain Toolbox.*

**Metabolite Maps and Estimates**

The metabolite maps of tNAA, Ins, and tCho are plotted in Figure 20 with the slice acquired with the 2D sequence used as reference for consistency check between the methodologies. A difference in coverage was observed between the modalities on the reference slice: on average on the entire datasets (slices acquired and rats), (73 ± 12)% of voxels passed the QC for 2D, while (76 ± 5)% and (58 ± 18)% were accepted for 3D and Multislice, respectively. Of note the coverage in the Multilisce datasets heavily decreased for the lower slices. Moreover, an increase of the CRLB was noted for Multislice, especially for tCho where the increase was significant in voxels in the middle of the brain leading to unreliable fitting.

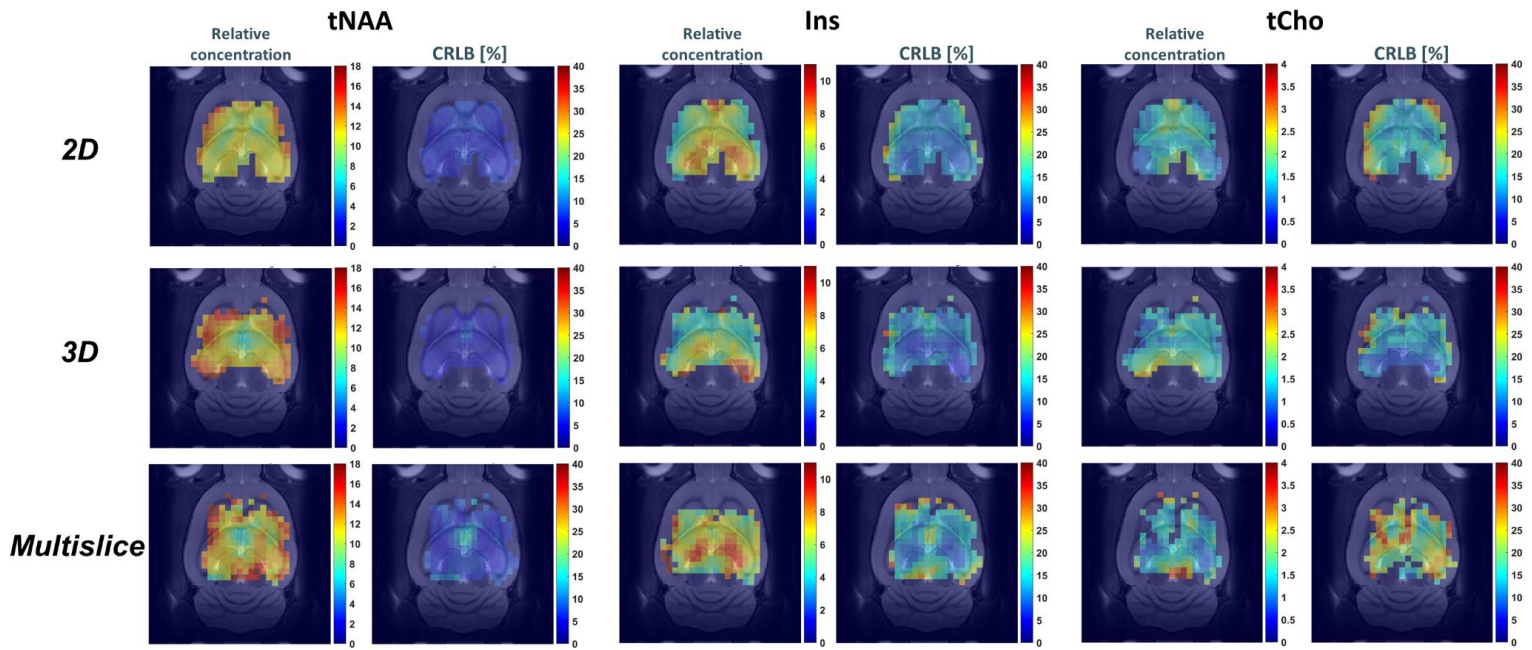

*Figure 20 Representative metabolite maps and CRLB maps of tNAA, tCho, and Ins, for 2D, 3D, and Multislice. The metabolite maps were superimposed to the corresponding anatomical image with the MRS4Brain Toolbox. The comparison between the 3 methodologies show a relatively comparable metabolite distribution for these 3 metabolites; with similar coverage (80% 2D, 75% 3D, 72% Multislice) for this specific rat. The metabolite maps scales correspond to LCModel outputs when referenced to tCr by setting its concentration to 8 mmol/kgww, while the CRLB scales correspond to a limit set at 40%.*

The reported mean relative concentration and CRLB values (for tNAA, Gln, Glu, Ins, Tau, and tCho) were measured through the shared common slice between the three methods (Figure 21). In the hippocampus region, both Ins and tNAA registered a coefficient of variation of 8% and 7% respectively, while the variation was below 6% for Gln, Glu, Tau, and tCho. The variations were more noteworthy on the striatum: a significant variation of 13% was observed for Tau and a smaller non-significant variation of 8% was measured for Ins. All other variations of the metabolite estimates were below 6%. While the difference in CRLBs between 2D and Multislice were not significant through the presented metabolites, the 3D CRLBs were consistently lower than the other two modalities (apart from tCho in the striatum), with statistically significant reduction for Glu, Gln, and tNAA. This can be in part explained by the general SNR increase, facilitating the LCModel fitting and reducing the uncertainty of the estimates.

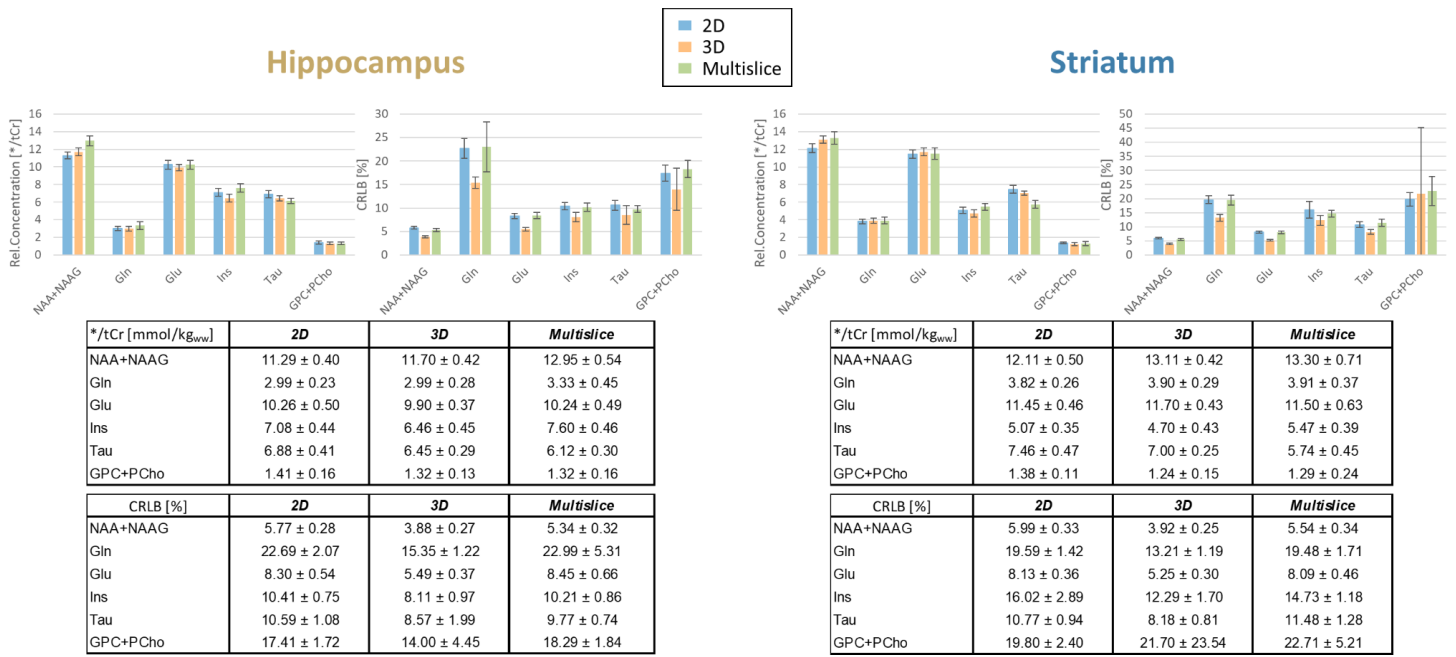

*Figure 21 Quantitative and CRLB assessment in each brain region (at 9.4T), for 2D, 3D, and multislice modalities (mean of the mean over 7 rats, per average). No significant differences were found in the hippocampus, while a significant difference was noted for Tau in the striatum (Multislice).*

# 4. Discussion

The present study demonstrated the application of CS for preclinical $^1$H-FID-MRSI acquisition at UHF (both 14.1 and 9.4T) to allow an acceleration of the acquisition from 13 minutes up to 3.25 minutes when using the highest AF (AF = 4), which is a novelty in the preclinical $^1$H-MRSI field. CS-FID-MRSI proved to be robust enough to provide similar results to standard FID-MRSI for all AFs with some decrease in the precision (i.e., increased SD) in the metabolite estimates and coverage. A comparable conclusion can be drawn out for the *Core* parameter of CS but a distinction between below and above 10% is needed, as there was a larger decrease of the map coverage for *Core* values below 10% due to an increase in parasitic lipid signals in the spectra and granularity of the metabolite maps.

With the adapted sets of parameters for CS, this technique was applied to further enhance the in-plane resolution by increasing the matrix size from 31 × 31 to both 47 × 47 and 63 × 63 without significantly impacting the metabolite distribution and mean estimates. Additionally, CS was combined with 3D FID-MRSI and Multislice FID-MRSI to acquire a larger volume of interest in the rodent brain in 29 minutes, while preserving the SNR advantages for the 3D and the lack of aliasing artifacts in the z-direction for the Multislice.

**CS Optimization on 2D FID-MRSI**

As one the main goals of this study was to provide an optimized CS protocol usable for higher resolution FID-MRSI at UHF in preclinical settings, the different CS acquisition parameters available (AF and *Core*) were explored to evaluate if changes occurred in the metabolite distribution compared to non-accelerated FID-MRSI. On average, the mean metabolite estimates were not significantly altered by any AF tested (Figure 13), though an increase of 50% of the SD was noted (stronger at *Core* size 10% and below combined with high AF, reaching almost more than a double increase), reflecting the loss of smoothness in the metabolite maps (Figures 8, 9, 10). One of the causes for this reduction in precision is related to the fact that random undersampling generates aliasing artifacts that are noise-like. Thus, the peaks of interest in the spectra need to be above this noise level for the reconstruction algorithm to recover them reliably. Even though the usage of UHF in this study had a positive effect on the CS reconstruction due to the additional SNR provided, especially for high AF, the granular appearance in the metabolite maps suggested that the vendor CS reconstruction might be suboptimal, requiring additional fine-tuning. Furthermore, this also translates in repercussions of the *k*-space sampling on the PSF of the acquisition. Indeed, when reaching higher AF values, the PSF profile tended to be similar to a non-filtered PSF (Supplementary Figure 7), which, as illustrated by Simicic et al.[37], could drastically modify the metabolite distribution.

Moreover, these changes in PSF profiles also imply an increase of lipid signals inside the brain and, consequently, a loss of coverage in agreement with the spatial aliasing artifact resulting from the lipid ring in CS reconstructions. It is worth highlighting that, in the current work, saturation bands were used to decrease the amount of lipid contamination at the acquisition level and consequently their impact during CS reconstruction. However, for low *Core* size, this approach was not sufficient to reliably reconstruct data without lipid contamination. The dependence of the lipid contamination with the AF using CS was reported in clinical FID-MRSI by Nassirpour et al.[31] using a *k*-space pattern with a relatively small *Core* size (4.5%). This echoes the results obtained when looking at the increase of the lobe amplitude with the AF (Figure 3) and with the *in-vivo* measurements.

The behavior of the PSF with the *Core* size was more complex as demonstrated in Figure 4, that the FWHM of the PSF decreased with the *Core* size and the lobe amplitude illustrated a non-linear trend. However, the lobe shape was more affected: a smaller *Core*. size ($\leq 10\%$) will tend to enlarge the first lobe to 6 voxels of distance, increasing the reach of the contamination (Supplementary Figure 3) A larger *Core* size will increase the lobe amplitude, but the reach was far lower (2 voxels of distance), and contribution was located closer to neighboring voxels than what was found in small cores (Supplementary Figure 8). Consequently, the ideal *Core* value was estimated to be between 10% to 20% of the sampling, regardless of the AF.

## Application of CS for 2D-High Resolution MRSI

The gains of acquisition time while using CS allowed for exploration of reduced nominal voxel size. Using an AF = 4, MRSI acquisitions with a nominal voxel of 0.52 µl ($47 \times 47$) and 0.28 µl ($63 \times 63$) were achieved, with reliably fitted spectra as well as metabolite distributions that were unaltered compared to the standard FID-MRSI acquisition (Figure 15). However, due to the very small voxel size when increasing the spatial resolution, averaging was used to compensate for the SNR loss. This had repercussions on the SNR maps (Figure 14A), as the $47 \times 47$ configuration showed an increase of the SNR in the hippocampus region, typically due to the 3 averages used. However, the normalized SNR showed no significant differences between each method, confirming no effects of CS on the overall SNR. Metabolite estimates and CRLBs (Figure 16) were preserved for all resolutions.

Of note, because of the high AF and the large *Core* size used, the FWHM of the PSF corresponded to 2.7 times the nominal voxel size (1.8 for non CS FID-MRSI acquisitions). This implied that the effective voxel size (*$31 \times 31$*:3.86 µL / *$47 \times 47$*:3.81 µL / *$63 \times 63$*:2.12 µL, reported in Supplementary Table 2) was reduced by 1.4% from $31 \times 31$ to $47 \times 47$ and by 45.1% from $31 \times 31$ to $63 \times 63$, which was significantly less than what one would expect from the 56.1% and 75.6% nominal voxel reduction. Thus, there is a clear incentive, when using CS, to acquire at much higher resolution to compensate for this inflation of the voxel size. In spite of that, the SNR found in the voxel will be reduced when acquiring with higher resolution, and to achieve acceptable or similar spectral quality as in non-accelerated FID-MRSI, averaging was used consequently increasing the acquisition time.

## Application of CS for 3D-MRSI

Another usage of the gained acquisition time explored in this study was the extension to 3D-MRSI. Due to the size limitations of the rat brain, and to allow a higher number of phase encoding in the z-direction, the voxel thickness was reduced from 2 mm to 1 mm. To cover almost the entirety of the brain in the coronal axis, the FOV was set at $24 \times 24 \times 9$ mm for a $31 \times 31 \times 9$ matrix size (nominal voxel size of $0.77 \times 0.77 \times 1$ mm or 0.59 µl). For comparison, the voxel size for 2D MRSI in the clinical field can reach up to $1.7 \times 1.7 \times 10$ mm (28.9 µl) for a $128 \times 128$ matrix[44]. Since the number of coronal slices acquired with 3D MRSI is relatively low, Multislice-MRSI was used as a viable alternative as it can avoid signal aliasing on the extremities of the FOV.

The intrinsic difference between the two types of acquisition lead to a *k*-space sampling that is unique in each method, as illustrated in Figure 1. The 2D sampling pattern selected (AF = 4; *Core* = 20%) was repeated for each slice of the Multislice configuration, as the 3rd dimension was acquired with the slice-selective pulse. This procedure is acceptable as the coronal slices were not sampled in the conventional manner with phase encoding and were, therefore, independent from the CS undersampling process. Accordingly, the PSF profile in the coronal direction was dependent on the excitation pulse response, as demonstrated in Figure 17, and can be much more efficiently optimized to avoid through-plane voxel bleeding. However, the usage of slice selection makes the Multislice more sensitive to chemical shift displacement error as opposed to 3D (~0.05 mm per ppm, 10% of the slice thickness). While the PSF FWHM in the coronal direction was larger for 3D compared to Multislice, the systematic displacement of the slice is avoided with the 3D modality, resulting in a well-positioned slice at the cost of a greater effective voxel size (1.6 times bigger for 3D compared to Multislice).

The CS parameters used for Multislice were, however, not applicable for 3D due to the core geometry and the low number of encoding steps in the coronal direction. The reduction of the *Core* size from 20% to 3% impacted the PSF profiles, as expected, from the parameters study conducted, resulting in smaller lobe amplitudes but with longer reach. The 2D-profile acquired from the coronal plane of the 3D acquisitions (Supplementary Figure 9) remained fairly acceptable with these parameters compared to what was observed for 2D-MRSI acquisitions in Supplementary Figure 8 (*Core* = 6%; stronger contributions from far-away lobes). One way to interpret this result is by considering the fact that the effective volume of the *Core* fully sampled with 3D is greater than what was sampled in 2D.

The adapted shimming procedure proved to be effective in achieving a reasonable brain coverage over the entire volume of interest. However, MAPSHIM leads to an increased linewidth over the 3D volume, as highlighted by the linewidth map of the water signal (Figure 18B), where an increase of the linewidth was visible from 2D to either 3D or Multislice, especially on the borders of the brain region. On the other hand, the MRSI coverage over Slice 3 through 7 remained fairly stable between 60 and 70%, with only the Multislice acquisition breaking below 50% at Slice 6 (Figure 19). So while the quality of the coverage was reduced over a singular slice, it was compensated by the overall coverage achieved throughout the entire volume.

The metabolite estimates in hippocampus and striatum of both Multislice and 3D were relatively faithful with the 2D acquisition on the comparative slice for tNAA, Gln, Glu, Ins and tCho (Figure 21). However, in the case of Tau, the concentration estimates in the striatum from the Multislice were lower than what was measured in both 2D and 3D methodologies. Furthermore, the Multislice metabolite maps seemed to have inconsistent coverage within the striatum region (Figure 18), leading to less voxels selected in this region. This was illustrated by an increase in SD of the mean CRLB of Tau, Gln and Glu and by an SNR map that shows a decrease in the right striatum (Figure 18A). This deterioration was not observed in the 2D acquisition. The 3D method bypasses this SNR issue due to its larger excitation volume. This directly increased the overall SNR of all the spectra acquired and facilitates the fitting procedures, offering a clear advantage over Multislice. The CRLBs of tNAA, Gln, Glu, Ins and Tau were significantly reduced when using 3D acquisition, with some major inconsistencies remaining for tCho.

Overall, 3D methodology provides a noteworthy augmentation of the SNR with similar estimates than what can be acquired with 2D FID-MRSI. One has to, however, account for the increase in effective voxel size and scan duration due to the additional phase encoding. As an alternative, Multislice provided the advantage of precise through-plane delimitation by using a slice-selective pulse, albeit with a SNR similar to what would be achievable with 2D and chemical shift displacement errors that have to be considered.

**Limitations and Future Steps**

While CS was successfully tested and applied in a variety of methods, there are aspects that need to be mentioned in its implementation and usage. The geometry of the core could benefit in being circular or spherical for 3D in CS, instead of following a square or a parallelepiped shape. Indeed, although the matrix dimensions used for 2D and 3D would not allow for a difference in *Core* when using a circular shape, the higher resolution acquisition, especially 63 × 63, would benefit from it as the PSF would be more isotropic. Different reconstruction could also be explored for CS datasets. One possibility is to use another regularization term in the form of a Total Variation[16,45,46], which could lead to improved reconstructed metabolic distribution. Reconstruction algorithms with a more adapted MRSI-focus such as the LR-TGV reconstruction[27] can offer different advantages such as a noise-reduction of the data, though one has to be attentive to possible biases for lower-concentrated metabolites in preclinical application[13].

Due to the rodent size constraints and the possibility of lipid contaminations from the sub-cranial regions, additional saturation bands were added compared to the 2D-FID-MRSI acquisition for the 3D and Multislice configurations and were positioned closer to the acquisition slab (Figure 1A). This has repercussions on the coverage of the first and last coronal slice of MRSI data and on the in-plane coverage. There are two ways to address this limitation: one could work on optimizing the saturation bands thickness and positioning to minimize unwanted crushing of metabolite signals on the VOI. This can be achieved by using Dixon fat-water separation method[46] during the imaging procedure to locate the main source of lipids and place the saturation bands accordingly. Alternatively, one could also explore thinner and fewer saturation bands combined with lipid suppression pre-processing steps such as the one presented by Bilgic et al.[45], although one has to be aware of baseline distortions caused by this process[27]. Acquiring data without any saturation bands as done previously for human CS FID-MRSI could also be an option[16,27,31]. However, the strong lipid contamination needs to be handled properly as it would lead to even stronger contaminations for CS due to spatial aliasing artifacts, as shown previously[31].

For the 3D MRSI acquisition, due to the higher SNR, an increase of the number of phase encoding steps in the coronal direction could be considered. This would also allow for a smaller effective voxel size and larger, more square shaped *Core* when using CS, leading to some improvement of the PSF profile in that specific direction. Additionally, metabolite map profiles in both sagittal and axial planes could be visualized more accurately with this increase in resolution. However, comparison with Multislice-MRSI in that case could not be deemed possible, as using a thinner slice thickness would reduce the signal acquired and, thus, the precision of the metabolite estimates would be affected. Moreover, this would also substantially increase the acquisition time (~3 minutes per additional slice with AF = 4), which was already shortened with CS to 29 minutes. Although it is possible to use higher AF values, this is not recommended as the lipid contamination can get more severe.[31] In this case, the better alternative would be the application of spatial-spectral encoding strategies which allow for higher AF without undersampling[10,44]. Moreover, it is possible to combine these encoding strategies with CS to achieve even higher spatial resolutions[10,47].

# 5.Conclusion

The present study illustrated the first implementation of a robust acceleration protocol for UHF $^1$H-FID-MRSI in the rat brain, reaching an AF up to 4 with non-significant alteration of the metabolite estimates and with a coverage loss averaging to 10% compared to the non-accelerated gold standard, mainly due to an increase of the CRLBs and a higher susceptibility to lipid contamination. This protocol was applied to increase the spatial resolution (from 0.77 × 0.77 to 0.38 × 0.38 mm of nominal voxel size) and to explore 3D MRSI acquisition with a reasonable acquisition time. While the gains in using CS for in-plane resolution increase were minor compared to the drawbacks related to the effective resolution, application of this acceleration technique was fairly successful for extension of the brain coverage using 3D and MultiSlice MRSI. Spectral quality metrics were preserved in each configuration, with a notable increase of SNR for 3D application. While useful for a first approach and application on preclinical [1]H-FID-MRSI acquisition, CS remains relatively limited due to potential lipid contamination and tempering of the effective resolution. Nevertheless, CS can be considered an effective tool with its ease of use and polyvalence, making it a good additional module to other acceleration strategies to gain additional temporal resolution for dynamic studies.

# Supplementary Materials

## 1. FID-MRSI Sequence Details

A Shinnar–Le Roux excitation pulse, adjusted to the Ernst angle of 52° and with an 8.4 kHz bandwidth, was applied for this sequence to achieve an acquisition delay AD = 1.3 ms. An acquisition bandwidth of 7.143 kHz with 1024 points was used, leading to a repetition time TR = 813 ms. The MRSI slice was done on the coronal plane with a 2 mm thickness and a field-of-view of 24 × 24 mm. The matrix size was set at 31 × 31, for a nominal voxel size of 0.77 × 0.77 × 2.00 mm (1.19 µL). VAPOR water suppression (hermite RF pulses, bandwidth: 600-660 Hz, flip angles 1 and 2: 84°/150°, last delay 22 or 26 ms, 614.7 ms duration) as well as 7 saturation bands (90° sech RF pulse, 1 ms, 20 kHz bandwidth) with thicknesses ranging from 4 to 10 mm were used.

## 2. CS Reconstructions

Both RAW and CS MRSI *k*-space data are stored in a matrix $S \in \mathbb{C}^{N_k \times N_t}$ before reconstruction and processing, where $N_k$ is the number of *k*-space points sampled and $N_t$ is the number of time points sampled. To express the data in the image space in a matrix $M \in \mathbb{C}^{N_r \times N_t}$, where $N_r$ is the overall number of spatial location, the Fourier transform operator is defined as a matrix $\mathscr{F} \in \mathbb{C}^{N_k \times N_r}$ where the element $\mathscr{F}_{i,j} = e^{ik_i \cdot r_j}$ is calculated from the scalar product between *k*-space point $k_i$ and position $r_j$. The inverse Fourier transform $\mathscr{F}^{-1}$ is defined as the conjugate transpose of $\mathscr{F}$ ($\mathscr{F}^{-1} = \mathscr{F}^{\dagger}$). Due to the changes in *k*-space sampling when performing CS, the Fourier operator is consistently computed before any of the reconstructions.

The reconstruction is based on the CS-MRI reconstruction algorithm described by Lustig et al.[30] and was applied online (before processing), with the Paravision 360 v3.3 and v3.5 interface. The principle of this method is to reduce the signal matrix $S$ to an image vector $y \in \mathbb{C}^{N_k \times 1}$ by taking each time point data individually: $y^{(i)} = S_{-,i}$ where $S_{-,i}$ is the i-th column of $S$. A sketch of the procedure can be seen in Supplementary Figure 2. A fast iterative shrinkage-thresholding algorithm (FISTA) is then used to minimize the following cost function :

$$argmin_{m^{(i)}} ||\mathscr{F}m^{(i)} - y^{(i)}||_2^2 + \lambda ||\Psi m^{(i)}||_1$$

where $m^{(i)} \in \mathbb{C}^{N_r \times 1}$ is the reconstructed image. A $\ell^1$-regularization is used with this reconstruction to promote sparsity of the solution[48]. $\Psi$ is a sparsifying operator and $\lambda$ is the parameter trade-off between the error ($||\mathscr{F}m^{(i)} - y^{(i)}||_2$) and the sparsity term. The minimization of the cost function is done for all $i$ and each $m^{(i)}$ is concatenated together to form the MRSI matrix $M$. For this study, the parameters used for the reconstruction were set with $\lambda = 10^{-3}$ and with a wavelet operator (HAAR wavelet, with a wavelet length of 2).

## 3. Homemade rat brain template and atlas-based segmentations

First, T2-weighted anatomical images of rat brains were aligned to generate a template using Advanced Normalization Tools (ANTs)[49]. Skull stripping of the generated template was then performed with the aid of Analysis of Functional NeuroImages (AFNI) tools[50] and FMRIB Software Library (FSL)[51]. Subsequently, the SIGMA template was registered to the generated rat brain template, yielding a transformation matrix. This matrix was used to transfer SIGMA rat brain segmentations to the template space. As a result, a homemade rat brain template with atlas-based segmentations was used in this study.

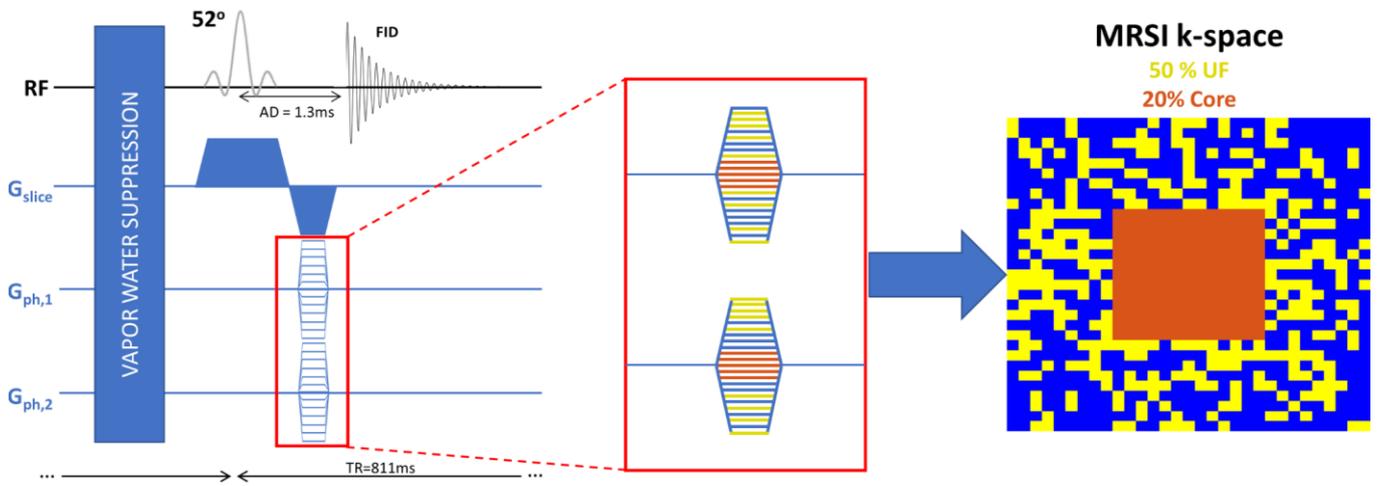

*Supplementary Figure 1: A schematic drawing of the $^1$H-FID-MRSI sequence with compress sensing (UF = 50%, Core = 20%) used for data acquisition. The resulting k-space mask shows an example of the distribution of the k-space points acquired with the above CS parameters.*

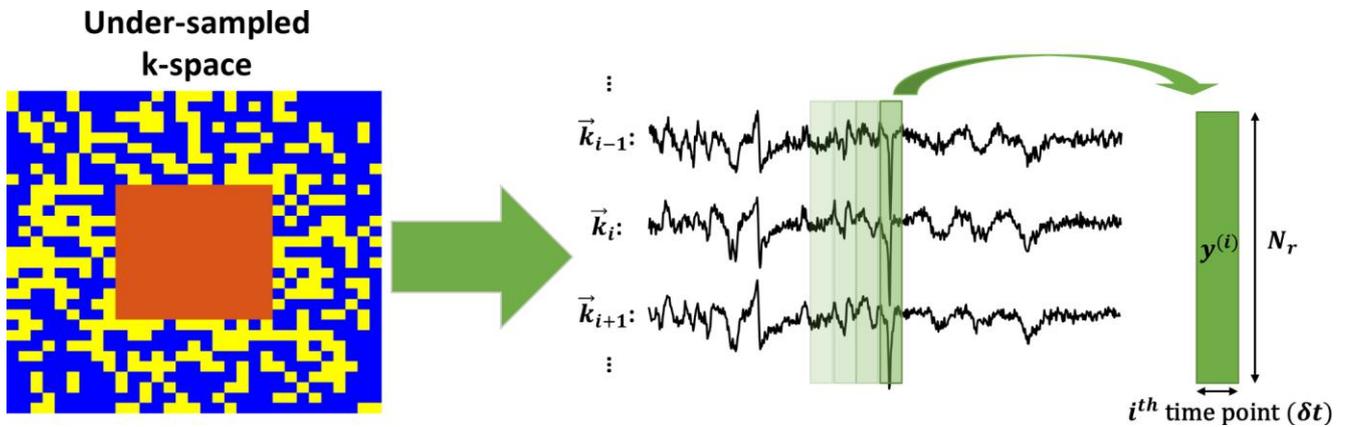

*Supplementary Figure 2: Sketch of the reconstruction processes: each spectrum acquired at different k-space points is concatenated together in a matrix. Each column is taken individually and used in the FISTA algorithm to obtain the representation in image space.*

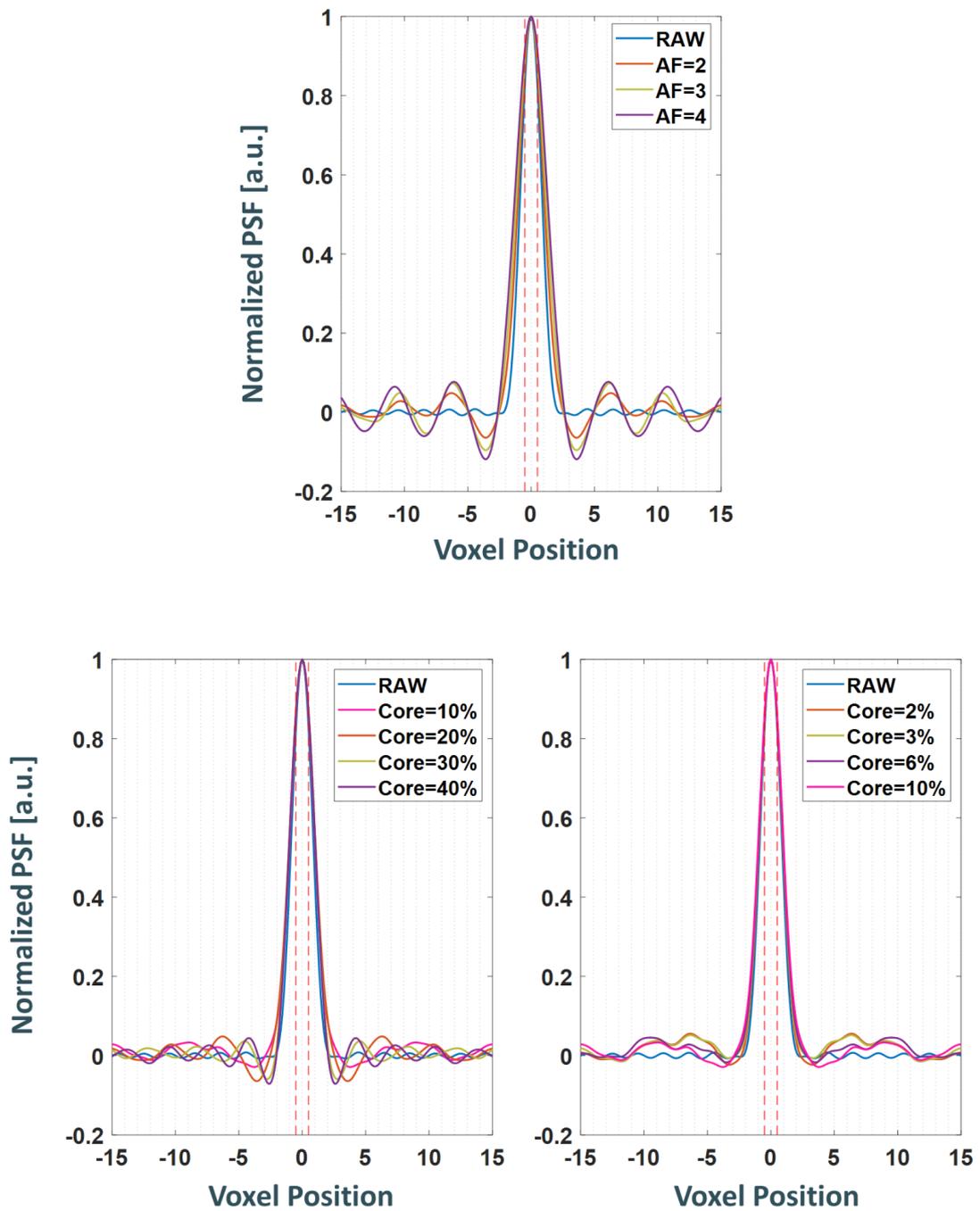

*Supplementary Figure 3: PSF of each configuration superimposed, for both AF and Core. The lobe amplitudes increase with the AF. For the Core parameter, one can see that the lobe amplitude decreases with the Core value, however the spread of the lobe is longer.*

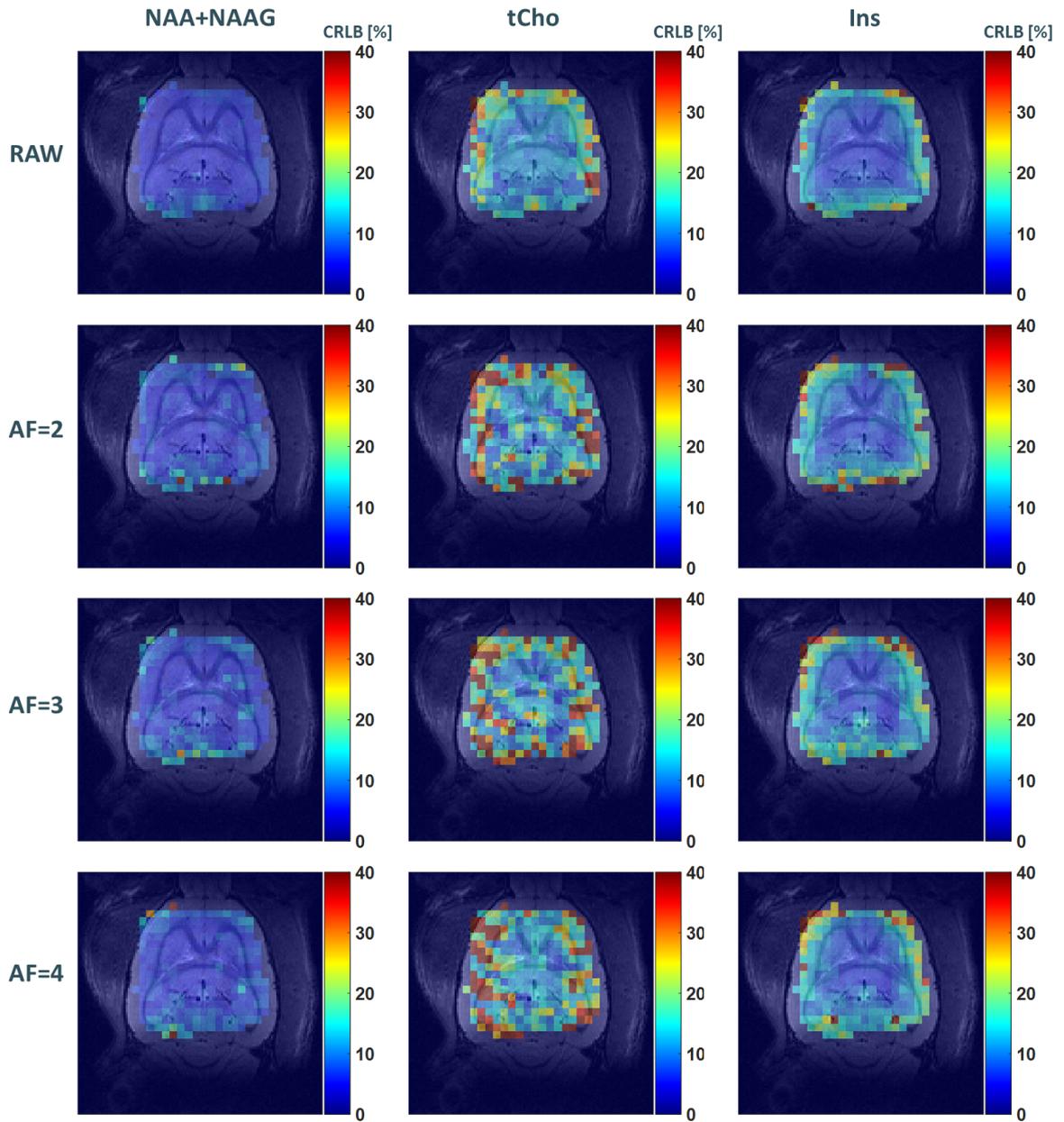

*Supplementary Figure 4: Representative CRLB maps of tNAA, tCho and Ins, for the RAW and different AFs configuration. The metabolite maps were superimposed to the corresponding anatomical image with the MRS4Brain Toolbox. The scales correspond to a CRLB limit set at 40%.*

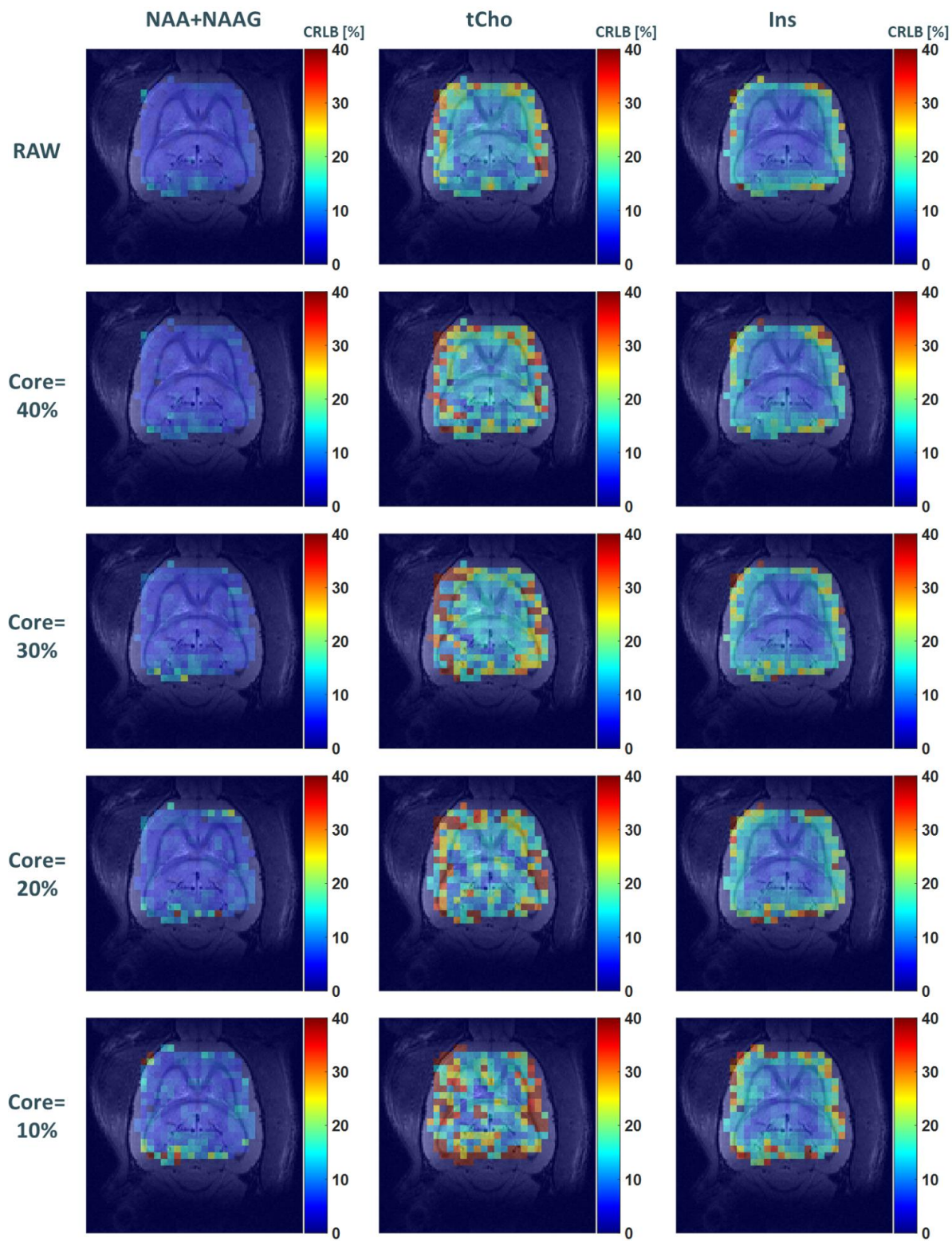

*Supplementary Figure 5: Representative CRLB maps of tNAA, tCho and Ins, for the RAW and different Core configuration (≥10%). The metabolite maps were superimposed to the corresponding anatomical image with the MRS4Brain Toolbox. The scales correspond to a CRLB limit set at 40%.*

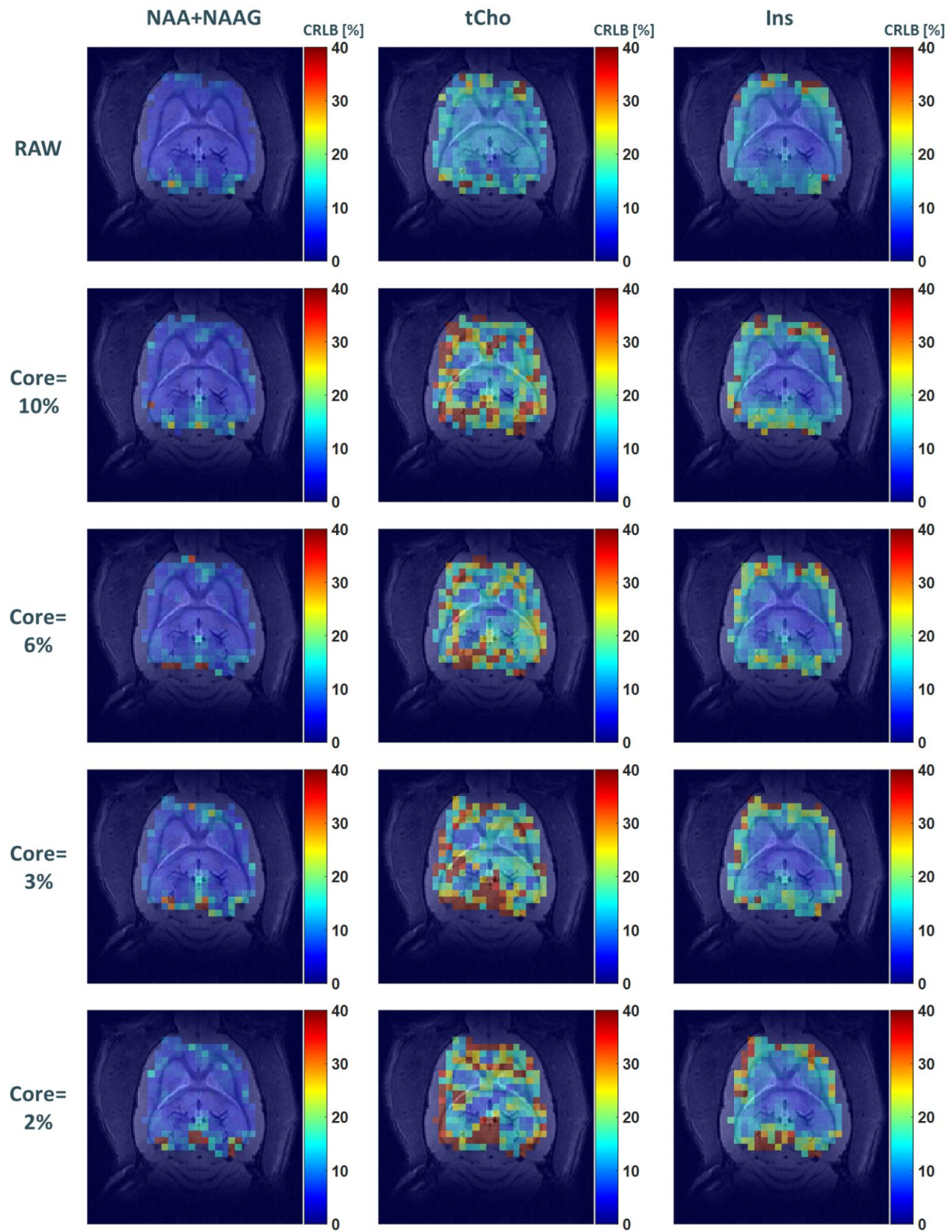

*Supplementary Figure 6: Representative CRLB maps of tNAA, tCho and Ins, for the RAW and different Core configuration (≤10%). The metabolite maps were superimposed to the corresponding anatomical image with the MRS4Brain Toolbox. The scales correspond to a CRLB limit set at 40%.*

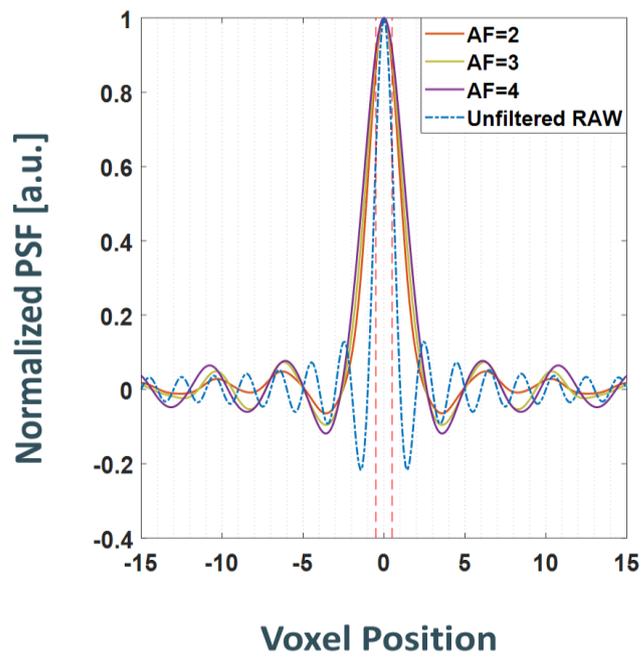

*Supplementary Figure 7: PSF of each AF configuration superimposed, compared to an unfiltered non-accelerated acquisition (unfiltered RAW, in dashed blue line). The side lobes tend to increase in amplitude and in width with the AF.*

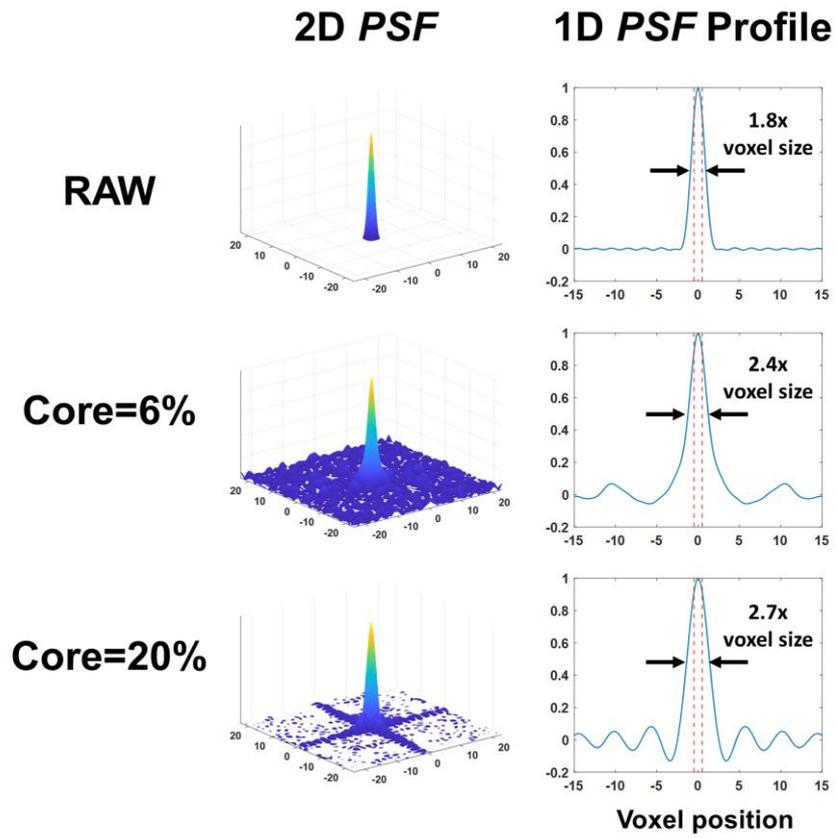

*Supplementary Figure 8: PSF profile (2D and 1D) for no CS, CS with AF = 4 and core sampled = 6% and CS with AF = 4 and core sampled = 20%. Core = 20% possess a larger FWHM value, but the 2D profile illustrates much less lobes around the central PSF peak compared to Core = 6%.*

## 2D PSF Profiles

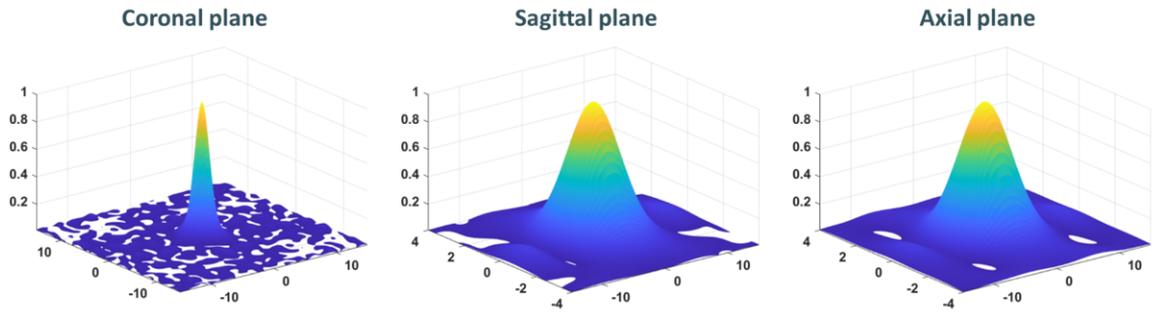

## k-space sampling

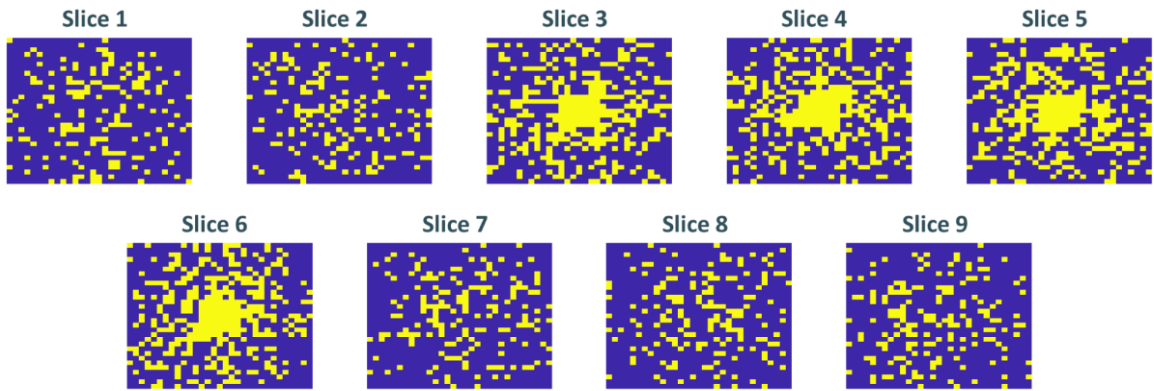

*Supplementary Figure 9: Point Spread Function (3D configuration) 2D profiles from each plane, and the k-space sampling using CS with AF = 4 and Core = 3%. The larger FWHM in the sagittal and axial plane are caused by the low number of phase encoding in the z-direction (9 phase encoding steps)*

**Supplementary Table 1:** Minimum reporting standards in MRS

## Hardware

| | |
|---|---|
| **Field strength** | 14.1T (CS Optimization) and 9.4T (High-Res, 3D, Multislice) |
| **Manufacturer** | Bruker |
| **model (software)** | Paravision 360 V3.3. (CS Optimization) and V3.5. (High-Res, 3D, Multislice) |
| **rf coil** | $^1$H-quadrature surface head coil (CS Optimzation) and $^1$H-quadrature volume-transmit coil and a cryogenic four-channel receive array (High-Res, 3D, Multislice). |
| **additional hardware** | N/A |

## Acquisition

| | |
|---|---|
| **pulse sequence** | FID-MRSI |
| **volume of interest (voi)** | Rodent: Brain |
| **nominal voi size** | 0.77 × 0.77 × 2 mm$^3$ (CS Optimization) <br><br> 0.51 × 0.51 × 2 mm$^3$ (47x47) <br><br> 0.38 × 0.38 × 2 mm$^3$ (63x63) <br><br> 0.77 × 0.77 × 1 mm$^3$ (3D and Multislice) |
| **repetition time TR and Acquisition delay AD** | TR = 813ms / AD = 1.3 ms (CS Optimization) <br><br> TR = 822ms / AD = 1.3 ms (High-Res, 3D) |
| | TR = 7398ms / AD = 1.3 ms (Multislice) |
| number of excitations per spectrum | 1 average (CS Optimization, 3D, Multislice)<br><br>3 averages (High-Res) |
| additional parameters | $24 \times 24 \times 2$ mm$^3$ FOV (CS Optimization, High-Res) / $24 \times 24 \times 9$ mm$^3$ FOV (3D, Multislice);<br><br>matrix size 31×31 and 47x47 and 63x63 (CS Optimization, High-Res) / 31x31x9 (3D) 31x31 in 9 slices (Multislice);<br><br>Cartesian *k*-space sampling, In-plane Hamming *k*-space filter in each dimension post processing;<br><br>Compressed Sensing acceleration: AF = {2,3,4} (only 4 for High-Res, 3D, Multislice) / Core Size fully sampled = {2,3,6,10,20,30,40} [%] (only 20 for High-Res and Multislice, only 6 for 3D);<br><br>Seven (CS Optimization) and twelve (High-Res) and thirteen (3D, Multislice) saturation slabs |
| water suppression method | VAPOR |
| shimming method | Bruker MAPSHIM, first in an ellipsoid covering the full brain and further in a volume of interest centered on the MRSI slice, with a thickness of 2 mm (CS Optimization, High-Res) and 6 mm (3D, Multislice).; < 30 Hz (CS Optimization), < 20 Hz (High-Res) and < 25 Hz (3D, Multislice) for H$_2$O resonance |
| triggering or motion correction | N/A |

## DATA ANALYSIS

| | |
|---|---|
| **Analysis Software** | LCmodel (Version 6.3-1N) |
| **Processing step deviating from reference** | Custom Basis-Set for AD = 1.3ms (for both 14.1T and 9.4T) <br><br> Control files provided with the *MRS4Brain Toolbox* |
| **output measure** | Ratios to total Creatine |
| **quantification reference** | Basis-set including: alanine, aspartate, ascorbate, creatine, phosphocreatine, γ-aminobutyrate, glutamine, glutamate, glycerophosphocholine, glutathione, glucose, inositol, N-acetylaspartate, N-acetylaspartylglutamate, phosphocholine, phosphoethanolamine, lactate, taurine simulated using NMR ScopeB. Macromolecules acquired in-vivo with double inversion recovery FID-MRSI. |

## Data quality

| | |
|---|---|
| **reported variables** | SNR (reference to NAA) and linewidths (reference to water) both reported <br><br> Global linewidths estimated by LCModel |
| **Data exclusion criteria** | LCModel SNR > 4 and <br><br> LCModel FWHM < 1.25*LCModel |
| **quality measures of post-processing model fitting** | CRLB < 30% |
| **sample spectrum** | Figure 2,9,10,11,18,23/ Supplementary Figure 2 |

| Acquisition parameters | | | |
|---|---|---|---|
| Parameters | *31x31* | *47x47* | *63x63* |
| Nb of Averages | 1 | 3 | 3 |
| Acceleration factor | 1 | 4 | 4 |
| Core fully sampled | 100% | 20% | 20% |
| **Acquisition time** | **13,2 min** | **22,7 min** | **40,8 min** |
| Filter Applied | Hamming | Hamming | Hamming |
| Field of View | 24x24 mm | 24x24 mm | 24x24 mm |
| Nominal Voxel Size | 0,77x0,77x2 mm | 0,51x0,51x2 mm | 0,38x0,38x2 mm |
| **Effective Voxel Size** | **1,39x1,39x2 mm** | **1,38x1,38x2 mm** | **1,03x1,03x2 mm** |

*Supplementary Table 2: Summary of the acquisition parameters of sets for the High Resolution study (the acquisition time and effective voxel size are highlighted in bold). Calculation of the effective voxel size is explained in section 2.3)*